\documentclass[]{aa}  
\usepackage[colorlinks, citecolor=blue]{hyperref}
\usepackage{color}
\usepackage{colortbl}
\usepackage{amsmath}
\usepackage{graphicx}
\usepackage{txfonts}
\usepackage{mathtools}
\usepackage{units}
\definecolor{Gray}{gray}{0.7}
\definecolor{lightGray}{gray}{0.85}
\usepackage{footnote}
\usepackage{rotating}

\begin{document}

\title{Prediction of astrometric microlensing events from  {\it Gaia} DR2 proper motions}
\author{J. Kl\"{u}ter\inst{1},
        U. Bastian\inst{1},           
        M. Demleitner\inst{1},
        J. Wambsganss\inst{1,}\inst{2}}

   \institute{ Zentrum f\"{u}r Astronomie der Universit\"{a}t Heidelberg, Astronomisches Rechen-Institut, M\"{o}nchhofstr. 12-14, 69120 Heidelberg, Germany\\
   \email{klueter@ari.uni-heidelberg.de}\and
   International Space Science Institute, Hallerstr. 6, 3012 Bern, Switzerland}
\authorrunning{J. Kl\"{u}ter et al.}
\titlerunning{Prediction of Astrometric Microlensing Events from {\it Gaia} DR2}

   \date{Received 28 July, 2018; accepted 30 September, 2018}
 
  \abstract
   {Astrometric gravitational microlensing is an excellent tool to determine the mass of stellar objects. Using precise astrometric measurements of the lensed position of a background source in combination with accurate predictions of the positions of the lens and the unlensed source it is possible to determine the mass of the lens with an accuracy of a few percent.}
   {Making use of the recently published {\it Gaia} Data Release 2 (DR2) catalogue, we want to predict astrometric microlensing events caused by foreground stars with high proper motion passing a background source in the coming decades.}
   {We selected roughly 148\,000 high-proper-motion stars from {\it Gaia} DR2 with \(\mu_{tot} > 150\,\mathrm{mas/yr}\) as potential lenses. We then searched for background sources close to their paths. Using the astrometric parameters of {\it Gaia} DR2, we calculated the future positions of source and lens. With a nested-intervals algorithm we determined the date and separation of the closest approach. Using  {\it Gaia} DR2 photometry we determined an approximate mass of the lens, which we used to calculate the expected microlensing effects. }
   {We predict 3914 microlensing events caused by 2875 different lenses between 2010 and 2065, with  expected shifts larger than \(0.1\,\mathrm{mas}\) between the lensed and unlensed positions of the source. Of those, 
513 events are expected to happen between 2014.5 - 2026.5 and might be measured by {\it Gaia}.
For 127 events we also expect a magnification between  \(1\,\mathrm{mmag}\) and \(3\,\mathrm{mag}\).}
  
{}


   \maketitle
%

\section{Introduction}
Gravitational lensing has become  a powerful tool to study galactic and extragalactic objects \citep{2006AnP...518...43W}. It is used for example to investigate the mass distributions of galaxies, to determine the Hubble constant,  to discover distant quasars, and to find extrasolar planets. Gravitational lensing describes the deflection and magnification of background sources by an intervening massive object \citep{1915SPAW...47..831E, 1936Sci....84..506E}. For stellar lenses (microlensing), two images of the source are created, a bright image close to the unlensed source position and a fainter image close to the lens.  Both images merge into a so-called Einstein ring when the source is perfectly aligned with the lens. The characteristic size of this ring is given by the Einstein radius
\begin{equation}
\theta_{E}=\sqrt{\frac{4GM_{L}}{c^{2}}\frac{D_{S}-D_{L}}{D_{L}D_{S}}},
\label{equation:theta_E}
\end{equation}
where \(M_{L}\) is the mass of the lens and \(D_{S}\), \(D_{L}\) are the distances between the observer and the source or the lens \citep{1924AN....221..329C,1936Sci....84..506E, 1986ApJ...301..503P}. This is the most important quantity since it sets the scale for all lensing effects. For close-by stellar lenses (within \(1\,\mathrm{kpc}\)) and distant sources, the Einstein radius is typically of the order of a few milliarcseconds. This is much smaller than the angular resolution of most of the currently available instruments. Due to the relative motion of source, lens, and observer,  magnification and image geometry change over time. 
Up to now, mostly  photometric magnification has been monitored and investigated by surveys such as the Optical Gravitational Lensing Experiment \citep[OGLE,][]{2003AcA....53..291U} or the Microlensing Observations in Astrophysics \citep[MOA,][]{2001MNRAS.327..868B} and has also led to the discovery of many exoplanets \citep[e.g.][]{2015AcA....65....1U}, whereas the astrometric shift of the source was detected for the first time only recently  \citep[][]{2017Sci...356.1046S,2018arXiv180701318Z}.

Astrometric microlensing provides the possibility to measure the mass of a single star with a precision of about one percent \citep{1995AcA....45..345P}.
Furthermore, astrometric microlensing events can be predicted from stars with a known proper motion. This is the aim of the present study. 
For the prediction of astrometric events, faint nearby stars with high proper motions are of particular interest. 
High proper motions are preferred because the covered sky area within a given time is larger, hence microlensing events are more likely. Nearby stars are preferred because their Einstein radius is larger and therefore the expected shift is also larger, and faint lenses are favourable since the
measurement of the source position is less contaminated by the lens brightness.

The first systematic search for astrometric microlensing events was done by \cite{2000ApJ...539..241S}.
They found 146 candidates between 2005 and 2015. 
\cite{2011A&A...536A..50P} predicted 1118 candidates between 2012-2019. However, most of those predictions were based on  erroneous proper motions in  some of the  catalogues used and only  49 events show reliable proper motions.  High-accuracy proper motions are essential to make precise predictions.
Today the  Gaia mission \citep{2016A&A...595A...1G} provides the best data for such studies. 
Using the TGAS data of the first data release \citep[Tycho-Gaia Astrometric Solution,][]{2016A&A...595A...4L}, \cite{2018MNRAS.478L..29M} predicted one event caused by  a white dwarf in 2019. 
With the second data release from Gaia \citep[{\it Gaia} DR2,][]{2018arXiv180409365G}, we also have precise parallaxes,  which are necessary to calculate the mass of a lens afterwards, as well as the proper motion of the background source. These big improvements in data quality and quantity made much more precise predictions possible.
Using {\it Gaia} DR2, we reported two ongoing microlensing events in 2018 \citep{2018arXiv180508023K}. 
Further, \cite{2018arXiv180510630B}  determine 76 microlensing events between 2014.5 and 2026.5, \cite{2018arXiv180511638M} predict 30 possible photometric microlensing events between 2015.5 and 2035.5,  
and \cite{2018arXiv180610003B} report the prediction of 2509 astrometric microlensing events until the year 2100.
In the present paper we present our method of how to use the {\it Gaia} DR2 proper motions and parallaxes to predict astrometric microlensing events in the coming decades. In Sect. \ref{chapter:microlensing} we explain the photometric and astrometric signatures of microlensing and describe how to determine the mass of the lens from the observation of the microlensing event.  In  Sect. \ref{chapter:method}, our method to find microlensing events is explained in detail. In Sect.  \ref{chapter:results} we present the events predicted by our search. Finally, we summarize our results and present conclusions in Sect. \ref{chapter:conclusion}.


\section{Basics of microlensing}
\label{chapter:microlensing}
\subsection{Photometric microlensing}
The magnification of a source due to the focusing of the light by an intervening lens is called photometric microlensing. 
The magnifications (\(A_{-}\),\,\(A_{+}\)) of the two images \((+)\),\,\((-)\) only depend on the dimensionless impact parameter \(\boldsymbol{u} = \boldsymbol{\Delta\theta}/\theta_{E} \), where \(\boldsymbol{\Delta\theta}\) is the unlensed angular separation between lens and source. When both images are merged, which is usually the case when photometric effects are measurable,  the total magnification can be determined via \citep{1986ApJ...301..503P}
\begin{equation}
A = A_{+} + A_{-}  =  \frac{ u^{2}+2}{u\sqrt{u^{2}+4}} 
,\end{equation}
where \(u = \lvert \boldsymbol{u} \rvert \).
For large impact parameters \((u\gg 1)\), it can be approximated by \citep{2000ApJ...534..213D}
\begin{equation}
A  \simeq 1+ \frac{2}{u^{4}}, 
 \label{equation:magnification}
\end{equation}
which shows a strong decline towards large separations. 
For bright, unresolved lenses,  the flux of the  lens \(f_{L}\) has to also be taken into account. Considering this, the measured  magnification is given by 
\begin{equation}
A_{lum} = \frac{f_{LS}+A}{f_{LS}+1}
,\end{equation}
where \(f_{LS}=f_{L}/f_{S}\) is the flux ratio between lens and (unmagnified) source star.  
In units of magnitude it is given by 
\begin{equation}
\Delta m = 2.5 \cdot\log_{10}\left({\frac{f_{LS}+A}{f_{LS}+1}}\right). 
\label{equation:approx_mag}
\end{equation}
Due to the strong decline with $u$, 
a measurable photometric magnification can only be observed when the impact parameter is small (i.e. on the order  of the Einstein radius or smaller). Therefore,  the timescale of a photometric microlensing event, given by the Einstein time,
 \begin{equation}
t_{E}=\frac{2\theta_{E}}{\mu_{rel}}
\end{equation} 
\citep{2012ARA&A..50..411G}, is quite short. Here,
\(\mu_{rel}\) is the absolute value of the relative proper motion between source and lens. Typical values for \(t_{E}\) are on the order of a few days or weeks. 

\subsection{Astrometric microlensing}

\begin{figure}
\includegraphics[width=9cm]{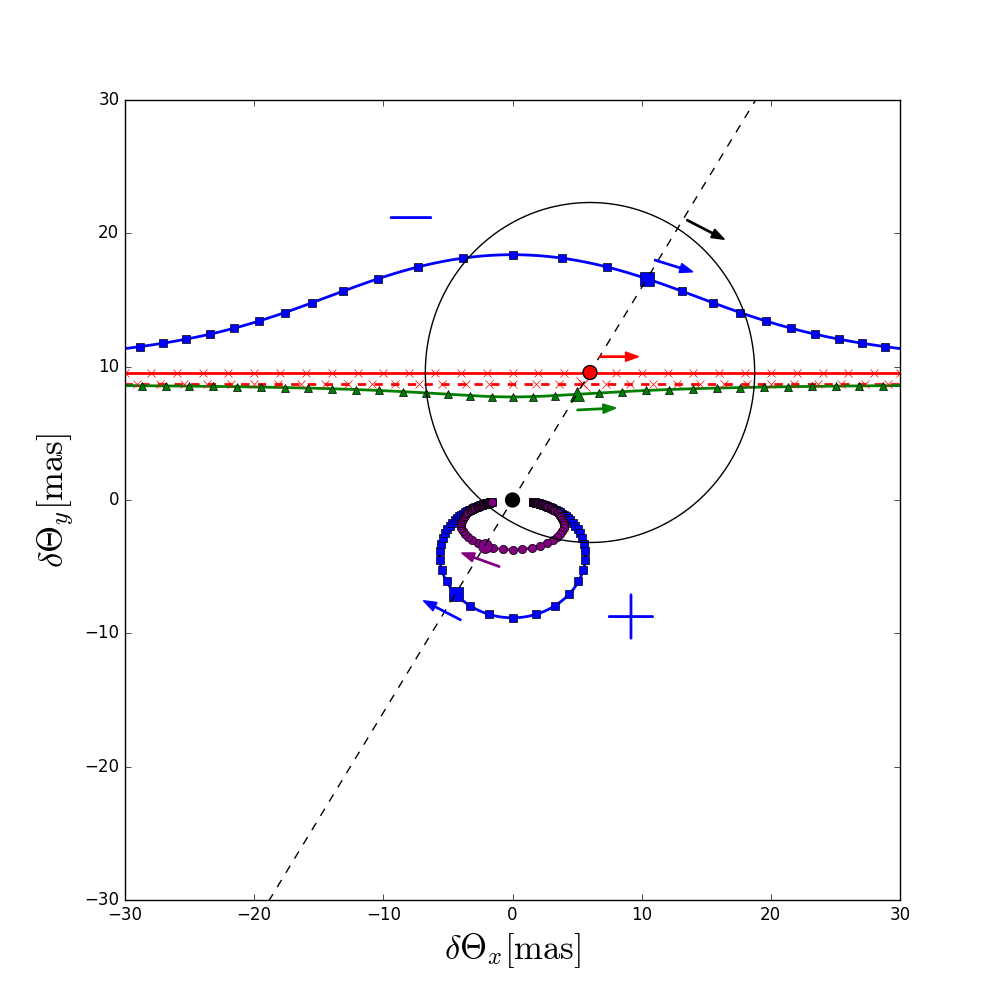}
\includegraphics[width=9cm]{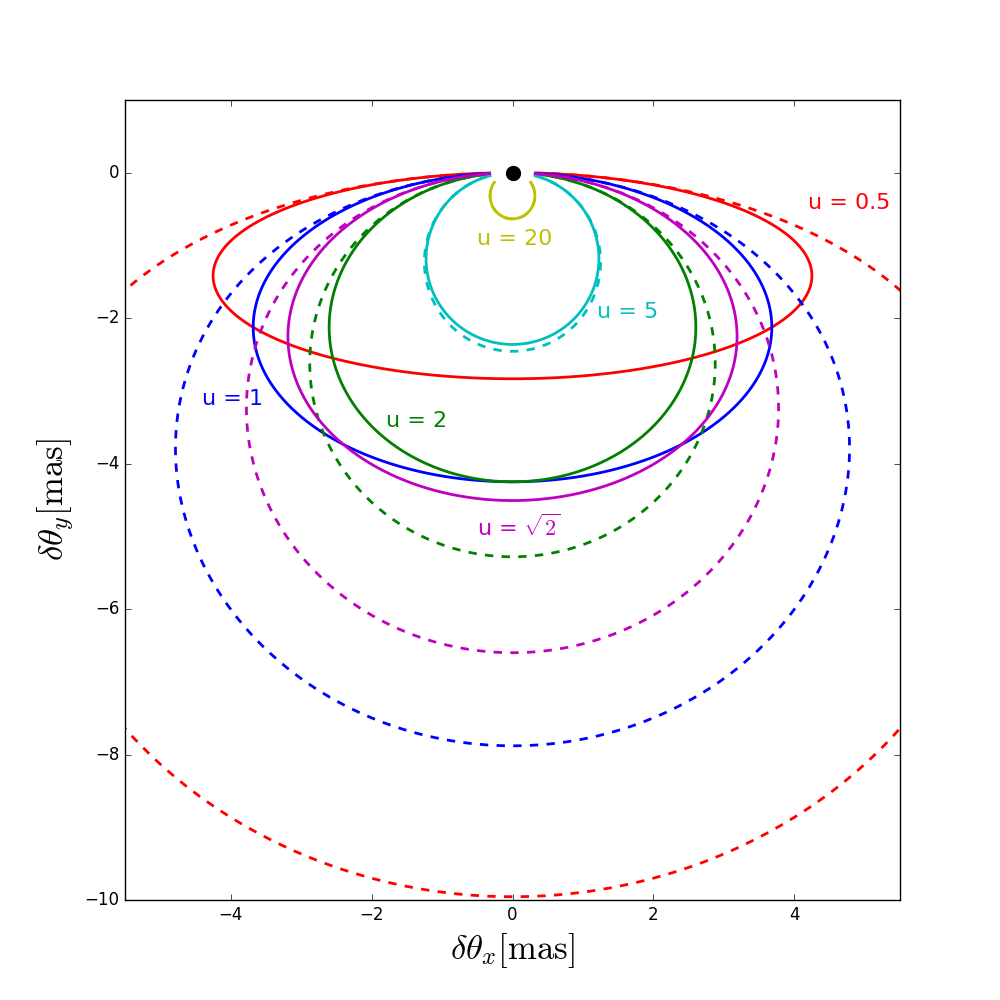}
\caption{ Top: Astrometric shift for an event with an Einstein radius of \(\theta_{E} = 12.75\)mas (black circle) and an impact parameter of u = 0.75. While the lens (red) passes a background star (black dot, fixed in origin) two images (blue) of the source are created due to gravitational lensing. This leads to a shift of the centre of light, shown in purple for a dark lens. In green, the centre of the combined light is shown for a flux ratio  of \(f_{LS} = 10\). The unlensed centre of the combined light is shown as a red dashed line. The black line connects the current positions of the snapshot. While the lens is moving in the direction of the red arrow, all other images are moving according to their individual arrows. The red, blue, and purple markers correspond to certain time steps \citep[after][]{2011A&A...536A..50P}. Bottom: Astrometric shift for different impact parameters. The black dot shows the fixed unlensed source position. The solid lines indicate the shift of the centre of light for a dark lens and the dashed lines indicate the shift of the brighter image. The maximum shift of the centre of light is reached at an angular distance of \(u =\sqrt{2}\) (purple) \citep{1998ApJ...494L..23P}, whereas the shift of the brightest image increases continuously with smaller distances.
}
\label{figure:shift}
\end{figure}

In astrometric microlensing, the change of the position of the background star is the signal of interest.
This is shown in the  top panel of  Fig.~\ref{figure:shift}.
The red line indicates a lens passing  a background source (black dot, fixed in the origin of the coordinate system). The two images created by the microlensing are shown in blue. The bright image \((+)\) is always close to the source and the faint image  \((-)\) is always close to the lens. Their positions relative to the lens can be described by \citep{1996ARA&A..34..419P} 
\begin{equation}
\boldsymbol{\theta_{\pm}} = \frac{ u \pm \sqrt{(u^{2}+4)}}{2} \cdot \frac{\boldsymbol{u}}{u} \cdot{\theta_{E}}
.\end{equation}
When the separation of the lensed images is too small to be resolved, only the position of the centre of light (purple line) can be measured. This can be expressed by 
\begin{equation}
\boldsymbol{\theta_c}= \frac{A_{+}\boldsymbol{\theta_{+}} +A_{-}\boldsymbol{\theta_{-}}}{A_{+}+A_{-}} =\frac{u^{2}+3}{u^{2}+2}\boldsymbol{u}\cdot{\theta_{E}}
\end{equation}
and the corresponding shift is given by
\begin{equation}
\delta\boldsymbol{\theta_{c}} = \frac{\boldsymbol{u}}{u^{2}+2} \cdot{\theta_{E}}.
\label{equation:shift}
\end{equation}
This is also a good approximation for the shift of the brightest image whenever \(u > 5\), since in this case the second image is negligibly faint. 
The astrometric effect reaches  a maximum value of
\(\delta\theta_{max} = 0.35 \theta_{E}\) at a separation of  \(u = \sqrt{2}\) (bottom panel of Fig. \ref{figure:shift}). For smaller separations, the effect will decrease \citep{1998ApJ...494L..23P} .

In the unresolved case, also luminous-lens effects usually have to be considered.  
The centre of light  of the combined system (green line in Fig. \ref{figure:shift}, top panel  ) can be expressed by \citep{1995A&A...294..287H,1995AJ....110.1427M}

\begin{equation}
\boldsymbol{\theta_{c,\,lum}} =  \frac{A_{+}\boldsymbol{\theta_{+}} +A_{-} \boldsymbol{\theta_{-}}}{A_{+}+A_{-}+f_{LS}} 
\end{equation}
and the shift between lensed and unlensed position can be determined via
\begin{equation}
\delta\boldsymbol{\theta_{c,\,lum}} = \frac{\boldsymbol{u}\cdot\theta_{E}}{1+f_{ls}}\,\frac{1+f_{LS}(u^{2}+3-u\sqrt{u^{2}+4})}{u^{2}+2+f_{LS}u\sqrt{u^{2}+4}}
.\end{equation}
For large impact parameters  (\(u\gg\sqrt{2}\)),  when the photometric effect  becomes negligible,  this simplifies to \citep{2000ApJ...534..213D}
\begin{equation}
        \delta\theta_{c,\,lum} \simeq \frac{\delta\theta_{c}}{1+f_{ls}} 
.\end{equation}
By using space telescopes like Gaia, or telescopes with adaptive optics, luminous-lens effects can be neglected for most of the astrometric microlensing events, since the separation between lens and source is larger than the angular resolution \citep[for Gaia \(FWHM = 103\,\mathrm{mas}\),][]{2016A&A...595A...3F}.
Such an instrument will measure the position of image (+).
The shift compared to the unlensed position of the source  can then be expressed by 
\begin{equation}
\delta\boldsymbol{\theta_{+}} = \frac{  \sqrt{(u^{2}+4)} - u}{2} \cdot \frac{\boldsymbol{u}}{u} \cdot{\theta_{E}}.
\end{equation}
For large impact parameters the shift is proportional to
\begin{equation}
\delta\theta_{+} \simeq \frac{\theta_{E}}{u}
\label{equation:approx_shift}.
\end{equation} 
Therefore, with increasing separation the astrometric shift drops much more slowly than the photometric magnification, that is, with \(1/u\) rather then with the fourth power (see Eq. (\ref{equation:magnification})). This results in a measurable effect at large separations and consequently in a much longer timescale during which an astrometric microlensing event can be observed \citep{1996AcA....46..291P,1996ApJ...470L.113M}. 
It can be described by \citep{2001PASJ...53..233H}  
\begin{equation} 
t_{aml} = t_{E}  \sqrt{\left(\frac{\theta_{E}}{\theta_{min}}\right)^{2} - u_{min}^{2}}
,\end{equation}
where \(\theta_{min}\) is the precision threshold of the used instrument. We consider a value of \(\theta_{min}= 0.1\, \mathrm{mas}\).  With such high-precision instruments, some events can be observed over a period of many months or even a few years. Hence astrometric microlensing can also be directly measured by high-precision, long-term surveys like Gaia, if the lensed stars are observed at a sufficient number of epochs suitably distributed in time.


\section{Prediction of microlensing events}
\label{chapter:method}
For the prediction of astrometric microlensing events, we use a method similar to \cite{2011A&A...536A..50P}.
The method consists of four steps:  1) Determine a list of high-proper-motion stars as potential lenses. 2) Find background sources close to their paths on the sky. 3) Forecast the exact position of source and lens stars from their current positions, proper motions, and parallaxes as well as determine the angular separation and epoch of the closest approach.  4) Calculate the expected microlensing effects, that is, the shifts of the background star positions. 

\subsection{List of high-proper-motion stars}

\begin{figure}
\includegraphics[width = 9 cm]{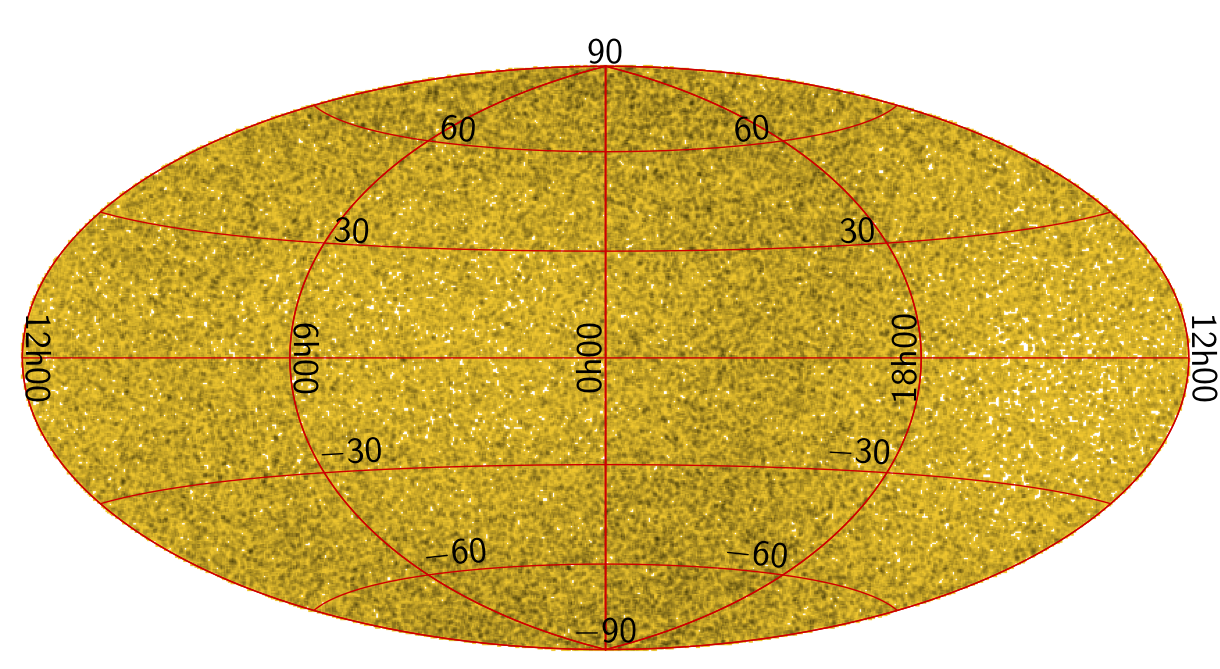}
\includegraphics[width = 9 cm]{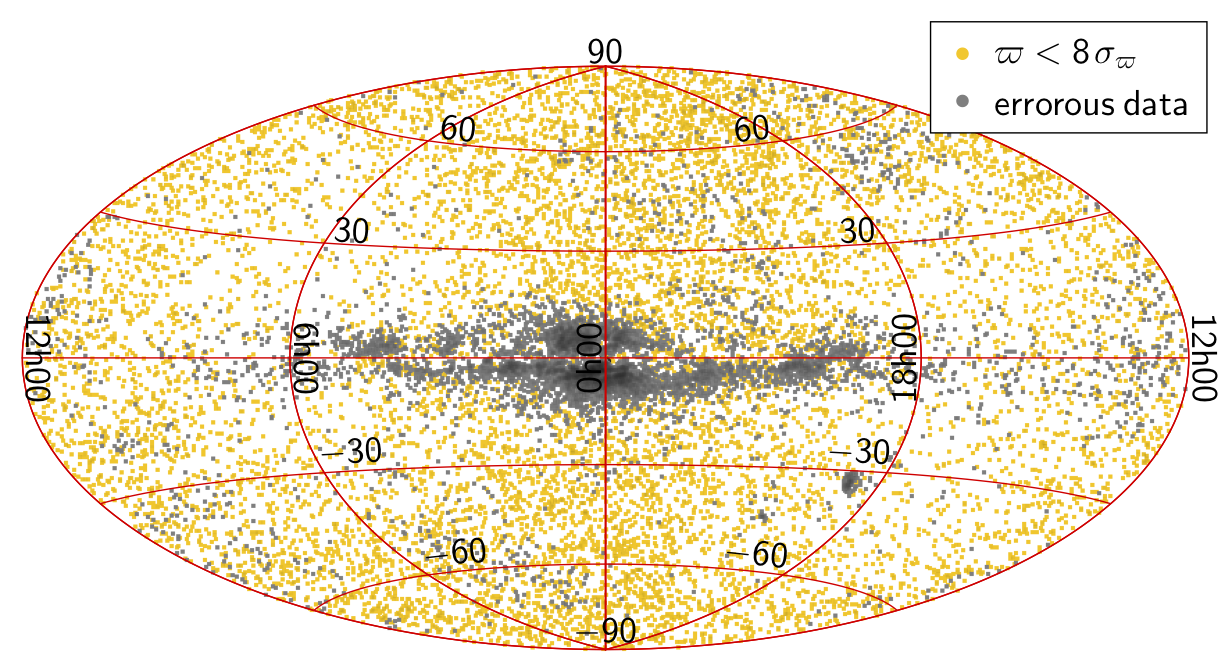}
\caption{Top: Aitoff projection all-sky map for the high-proper-motion stars  with \(\mu_{tot} > 150\, \mathrm{mas/yr}\) in galactic coordinates. The small under-densities at (\(4h,\,0^{\circ}\)) and (\(16h, \,0^{\circ}\)) are caused by the solar apex.
Bottom: All excluded objects. The yellow dots show the sources with non-significant parallaxes, the grey dots indicate the erroneous data. }
\label{fig:aitoff}
\end{figure}

\begin{figure}
\includegraphics[width = 9 cm]{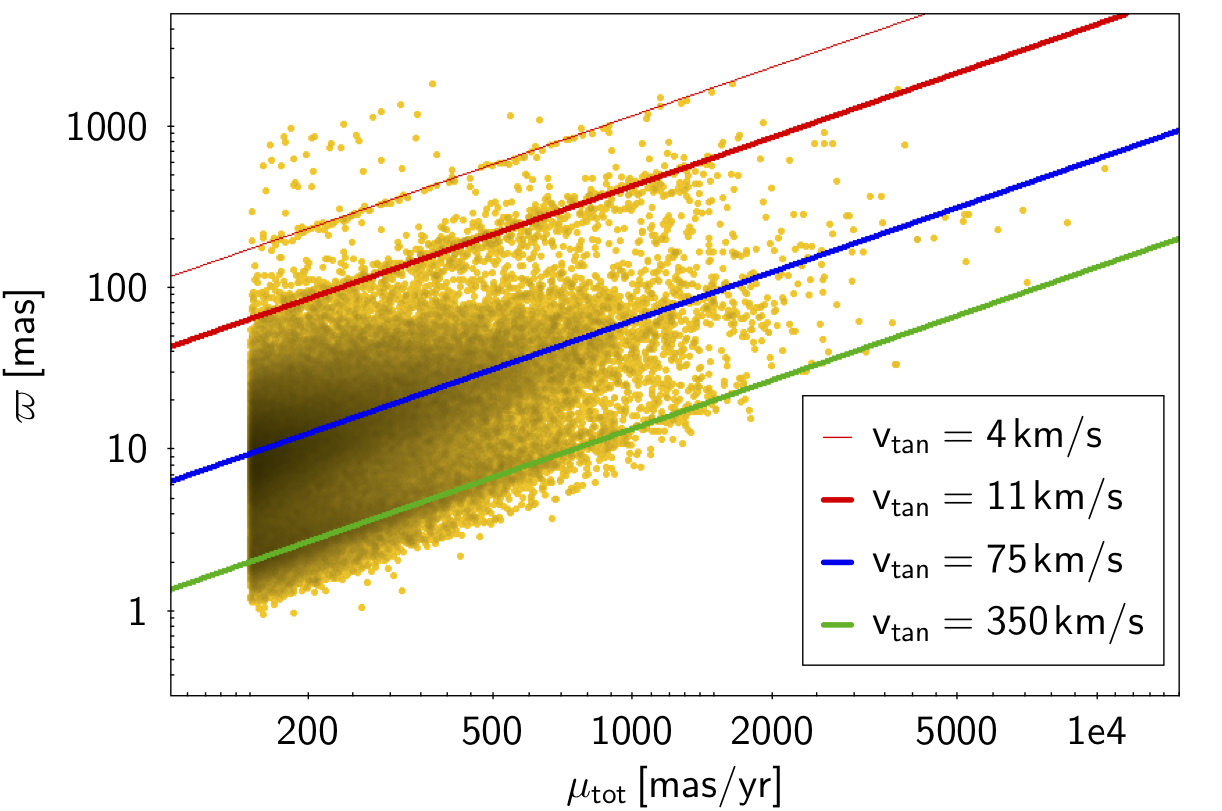}
\includegraphics[width = 9 cm]{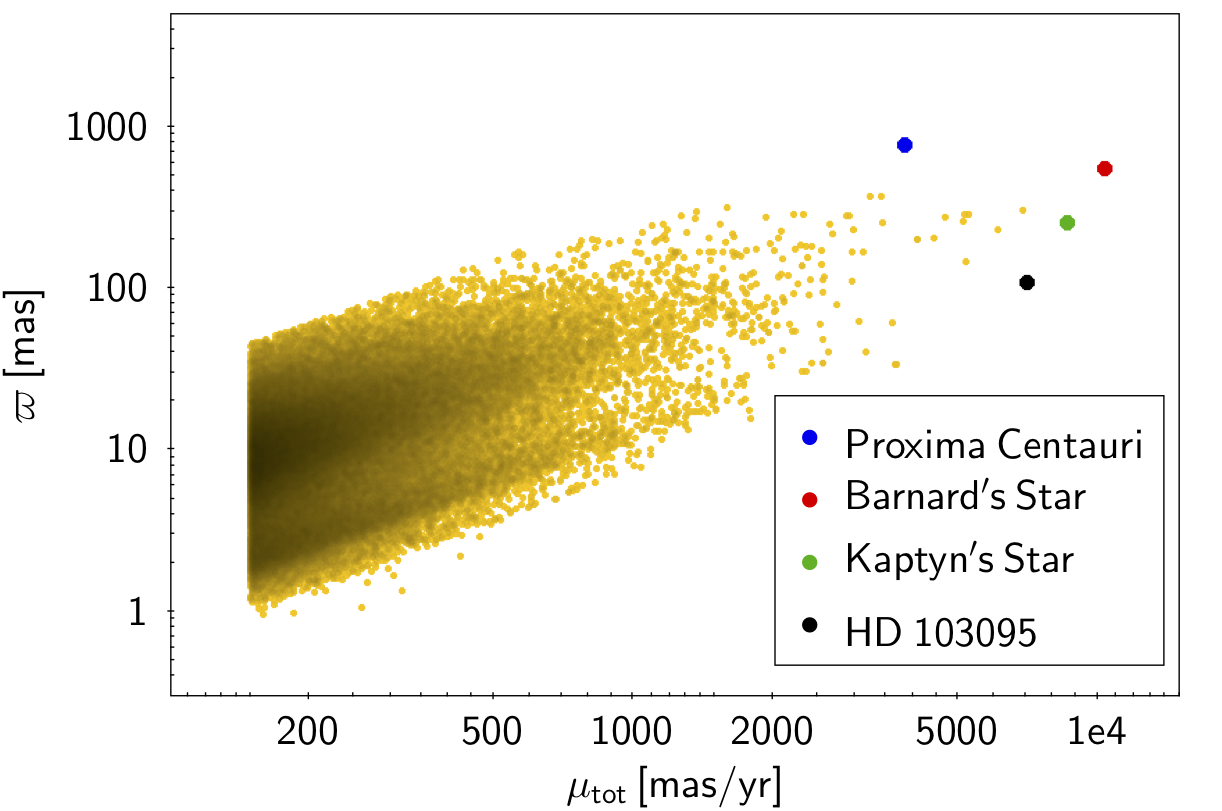}
\caption{Top: Proper motions \((\mu_{tot})\) and parallaxes \((\varpi)\) for all sources with significant parallaxes. The green line indicates the population of halo stars, the blue line indicates the population of disc stars, and the red lines indicate two sharp populations of obviously erroneous objects. Bottom: Proper motions and parallaxes for the cleaned sample. The isolated points with very high proper motion correspond to real stars, for example  Proxima Cen (blue), Barnard's star (red),  Kapteyn's star (green), and  HD 103095 (black).}
\label{fig:px_vs_pm}
\end{figure}

\begin{figure}
\includegraphics[width = 9 cm]{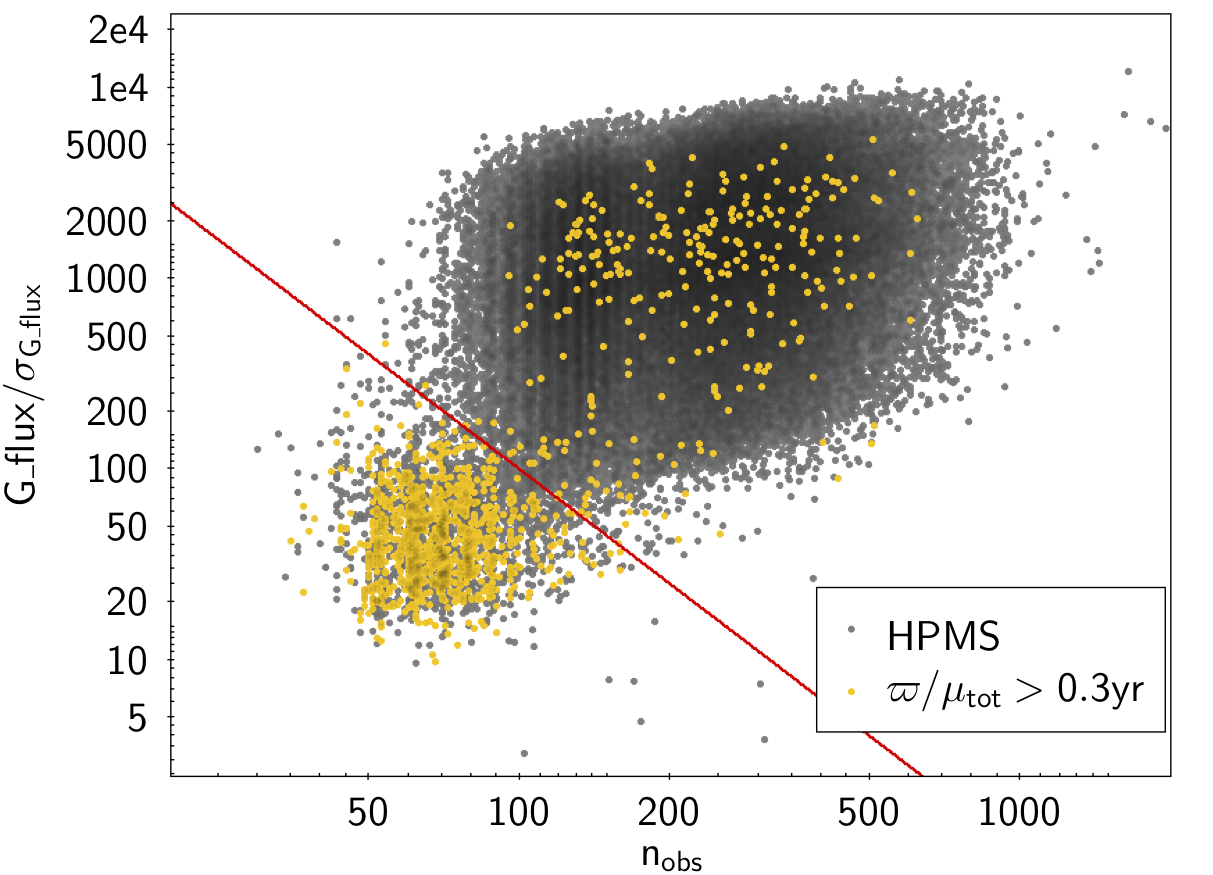}
\caption{Number of photometric observations by Gaia (\(n_{obs}\))  and  significance of the G flux (\(G\_flux/\sigma_{G\_flux}\)) for all high-proper-motions stars with significant parallaxes. The yellow dots indicate the sources with \(\varpi/\mu_{tot} > 0.3\,\mathrm{yr}\). The red line indicates our used limit. 
The excluded lenses  right above this limit are most likely real objects.} 
\label{fig:n_obs}
\end{figure}

 Due to its unprecedented accuracy, the {\it Gaia} DR2 provides the ideal catalogue for this task. {\it Gaia} DR2 contains roughly 170\,000 sources with proper motions \(\mu_{tot} = \sqrt{\mu_{\alpha^{\star}}^{2}+\mu_\delta^{2}}\) larger than \(150\,mas/yr\). As the Gaia Consortium has mentioned \citep{2018arXiv180409366L}, DR2 contains a small proportion of erroneous astrometric solutions, most noticeably a set of unrealistically high proper motions or parallaxes. To clean up our target list, we therefore first neglect all sources with insignificant parallaxes \((\varpi < 8  \sigma_{\varpi})\). This and all other quality  cuts used by us are shown in Table \ref{tab:cuts}. 
Figure \ref{fig:px_vs_pm} shows the absolute values of the proper motions and the parallaxes of the remaining high-proper-motion stars.
Four different  populations are clearly visible. The two lower ones are interpreted as the real populations of halo stars with a typical tangential velocity of \(v_{tan} \sim 350\,\mathrm{km/s}\) (green line), and disk stars (\(v_{tan} \sim  75 km/s\)) (blue line), whereas the two upper populations (red lines) are 
incorrect data, since such stars do not exist --- at least not in such numbers and at distances of 10\,pc or smaller. Why the faulty {\it Gaia} DR2 data show such sharp relations between parallax and proper motions is not yet known  (private communication from the Gaia astrometry group).   
 To exclude those faulty Gaia data, we neglect all stars with \(\varpi/\mu_{tot} > 0.3\,\mathrm{yr}\). These suspicious data are also well separated in Fig. \ref{fig:n_obs}, where the significance of the Gaia G flux  (\({G\_flux}/\sigma_{G\_flux}\))  is plotted against  the number  of photometric observations by Gaia (\(n_{obs}\)). Hence we exclude all sources with  \(n_{obs}^{2}\cdot G\_flux/\sigma_{G\_flux}< 10^{6}\) (i.e below-left of the red line).
Our final list contains \(\sim 148\,000\) high-proper-motion stars, which are potential lenses. 
As expected, these nearby objects are quite evenly distributed over the sky (Fig. \ref{fig:aitoff}, top panel), whereas the rejected objects mainly cluster towards the Galactic disc/bulge or the Magellanic clouds (bottom panel). 


\subsection{Background stars}
\begin{figure}
\includegraphics[width=9cm, ]{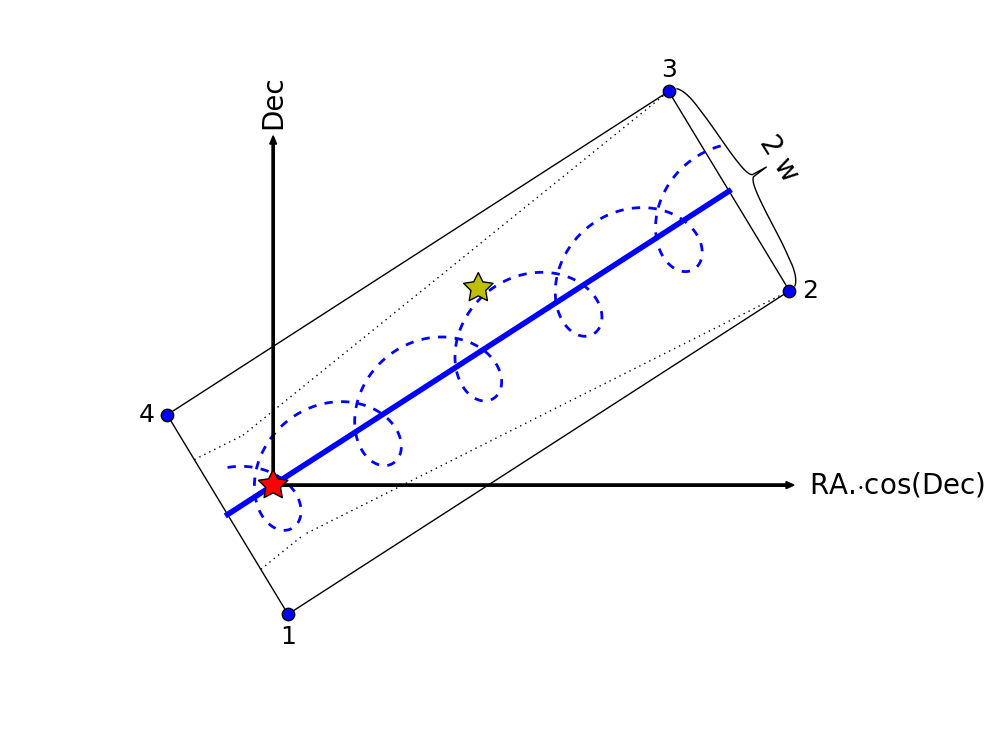}
\caption{Illustration of the window used in the search for background stars. The thick solid blue line indicates the proper motion of the lens (red star), and  the origin is set to the J2015.5 position of the lens.  The blue dashed line  indicates the real motion, which includes the parallax (only five years are shown). When a background star (yellow star)  is within the black box, it is considered as candidate. For the plot, the box is defined by the position of the lens in J2010.0 and J2015.5 and a half width of 7 arcsec. To account for the proper motion of the background source, the widening shape (dotted black line) would be more physically accurate.}
\label{figure:window}
\end{figure}

For each of the roughly 148\,000 remaining  high-proper-motion stars, we searched for background sources close to their paths.
For this, we defined a box by using the position of the source at the epochs J2010.0 and J2065.5 with a half-width \(w = 7''\) perpendicular to the direction of the proper motion.
This box  is illustrated in Fig.~\ref{figure:window}.
The large box width is mainly adopted  to account for potential motions of background sources. A widening shape would be more physically accurate, however, for simplicity we used the rectangular shape. 
The combination of a high-proper-motion foreground lens and a background source within the defined box  is called a ``candidate''. In the following, the source parameters are labelled with the prefix ``\text{Sou}\_''.

We considered all {\it Gaia} DR2 sources, without a significantly negative parallax  (\(\text{Sou}\_ \varpi > -3 \cdot \text{Sou}\_ \sigma _{\varpi} -0.029\,\mathrm{mas}\)) and with a standard error in the J2015.5 position below 10 mas \(\left(\sqrt{\text{Sou\_}\sigma_{ra}^{2}+\text{Sou\_}\sigma_{dec}^{2}} < 10\,\mathrm{mas}\right)\)  as potential background sources. 
For sources with non-significant negative parallaxes (\(-0.029\,\mathrm{mas} > \text{Sou}\_\varpi + 3 \cdot \text{Sou}\_\sigma_{\varpi}\)) or without parallax in {\it Gaia} DR2, we assumed a value of  \(\text{Sou}\_\varpi  = -0.029 \,\mathrm{mas}\). This is the zero-point of Gaia's parallaxes, as determined from a sample of known quasars \citep{2018arXiv180409376L}. By using this value, we were able to correct for this systematic error. 
For background sources that have only a two-parameter astrometric solution,  we assumed a standard error in the proper motion  of \(\text{Sou}\_\sigma_{\mu_{ra,dec}} = 10\,\mathrm{mas/yr}\), and a parallax error of \(\text{Sou}\_\sigma_{\varpi} =2\,\mathrm{mas}\).
Roughly 90\% of the five-parameter background sources have proper-motions and parallaxes below this value.

To avoid binary stars and co-moving stars in our candidate list, we exclude pairs with common proper motion, that is,
\begin{equation}
\lvert \vec{\mu}_{tot} - \text{Sou}\_\vec{\mu}_{tot}\rvert\,  < 0.7\cdot \lvert  \vec{\mu}_{tot} \rvert 
.\end{equation}
These criteria can only be used if the proper motion of the source is given in {\it Gaia} DR2.
This is not the case for roughly  \(25\%\) of our events.   
Hence, these events have to be treated carefully, especially when the estimated date of closest approach is close to J2015.5.
Nevertheless, most of them are expected to be real events.

Further, we exclude candidates where the parallax of the source is larger than the parallax of the lens (\(\text{Sou}\_\varpi > \varpi\)), to avoid negative Einstein radii.
We do not make a stronger cut for the parallax at this point since comparable parallaxes will lead to small Einstein radii anyway and hence to small astrometric shifts. 

\subsection{Position forecast and determination of the closest approach} 

For about  68\,000 candidates, we searched  for the closest approach by calculating the positions of source and lens from {\it Gaia} DR2 positions, proper motions, and parallaxes.
Since we are interested in the global minima and the periodic motion of the Earth may cause many local minima, we first neglected the Earth's motion to calculate an approximate distance and time of the closest approach, using a nested-intervals algorithm.
If the expected shift according to Eq. (\ref{equation:approx_shift}) for the approximate distance is larger than \(0.03\,\mathrm{mas}\), the exact value is calculated by including the parallax.
In order to account for the multiple minima, we searched for all local minima within \(\pm1\,\mathrm{year}\) around the approximate time with intervals of roughly four weeks. 
It is possible that we considered two really  close minima as one. However, the distances and dates of both minima should then be very similar. 
For all minima found, we determined the minimum separations and the epoch of the closest approaches, again using the nested-intervals algorithm. By comparing these values we selected the global minima. 

Since Gaia and the future James Web Space Telescope (JWST) are located at the Lagrange point~L2, we repeated our study with a 1\%\ larger parallax to take account of  the larger heliocentric orbit at L2. As expected, the effects only differ when the smallest separation is small compared to the parallax. 

\subsection{Approximate mass and Einstein radius}
\begin{figure}
\includegraphics[width=9cm ]{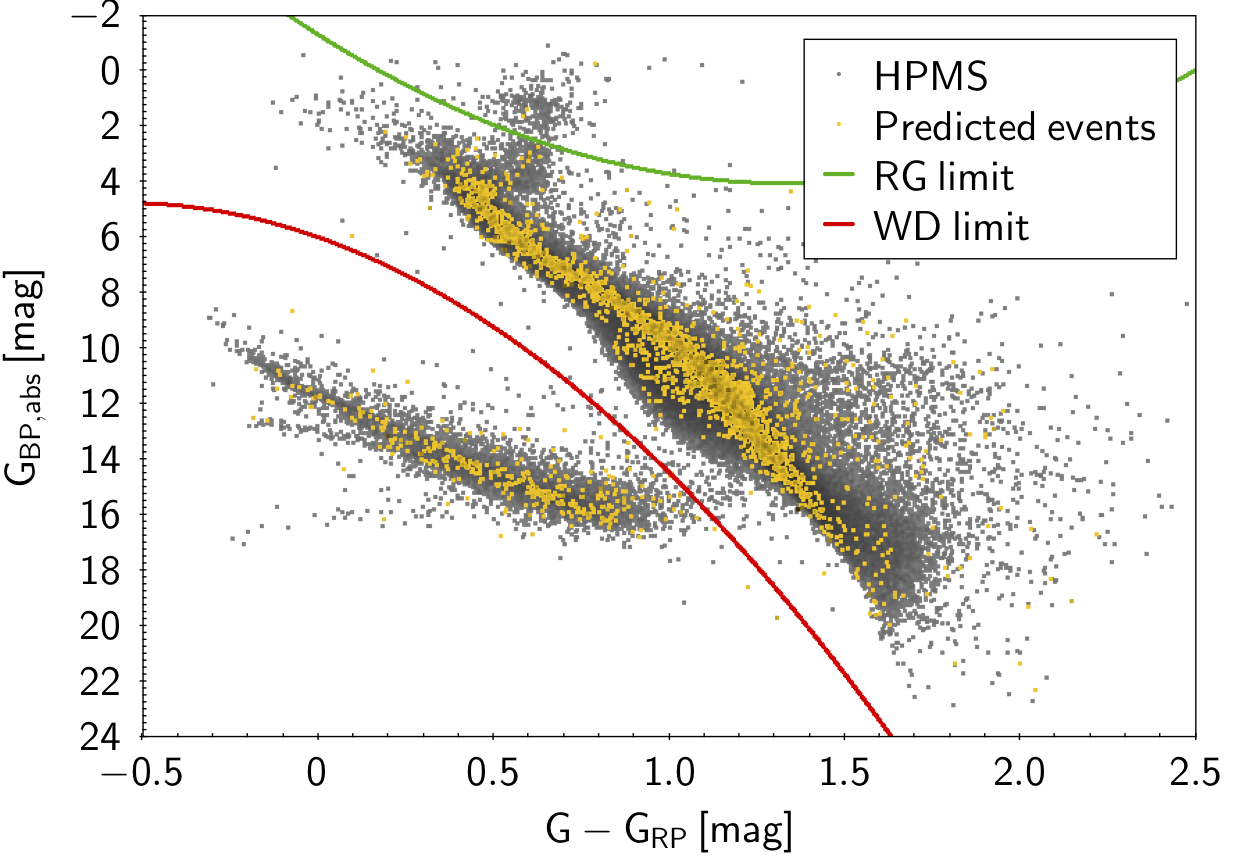}
\caption{Colour-magnitude diagram of all potential lenses with full {\it Gaia} DR2 photometry \(G\), \(G_{RP}\) , and \(G_{Bp}\)
The yellow dots indicate the lenses of the predicted events. All stars above the green line are considered as red giants and all sources below the red line as white dwarfs.}
\label{figure:cmd}
\end{figure}

In order to get a realistic value for the expected astrometric shifts of our candidates, we derived a rough approximation for the mass of each lens in the following way. 
First, we divided our candidates into three categories --- white dwarfs (WD), main sequence stars (MS), and red giants (RG) --- by using the following cuts in  colour-magnitude space (see Fig. \ref{figure:cmd}): 
\begin{equation}
\begin{aligned}
WD:&\qquad G_{BP, abs} \ge 4\cdot (G-G_{RP})^{2}+4.5\cdot(G - G_{RP}) +6\\
RG:&\qquad G_{BP, abs} \le -3 \cdot (G-G_{RP})^{2}+8 \cdot (G-G_{RP})  - 1.3
\end{aligned}
\label{colorcuts}
.\end{equation}
Here, \(G_{BP,abs}\) represents the absolute magnitude determined via the distance modulus.
Lenses without \(G_{BP}\) and \(G_{RP}\) magnitudes are assumed to be main-sequence stars. 
For white dwarfs and red giants, we used typical  masses  of  \(M_{WD} = (0.65\pm 0.15) \,M_{\odot}\) and  \(M_{RG} = (1.0\pm 0.5 \,M_{\odot})\), respectively, where the indicated uncertainties are used for the error calculus further below. 

\begin{figure}
\includegraphics[width=9cm ]{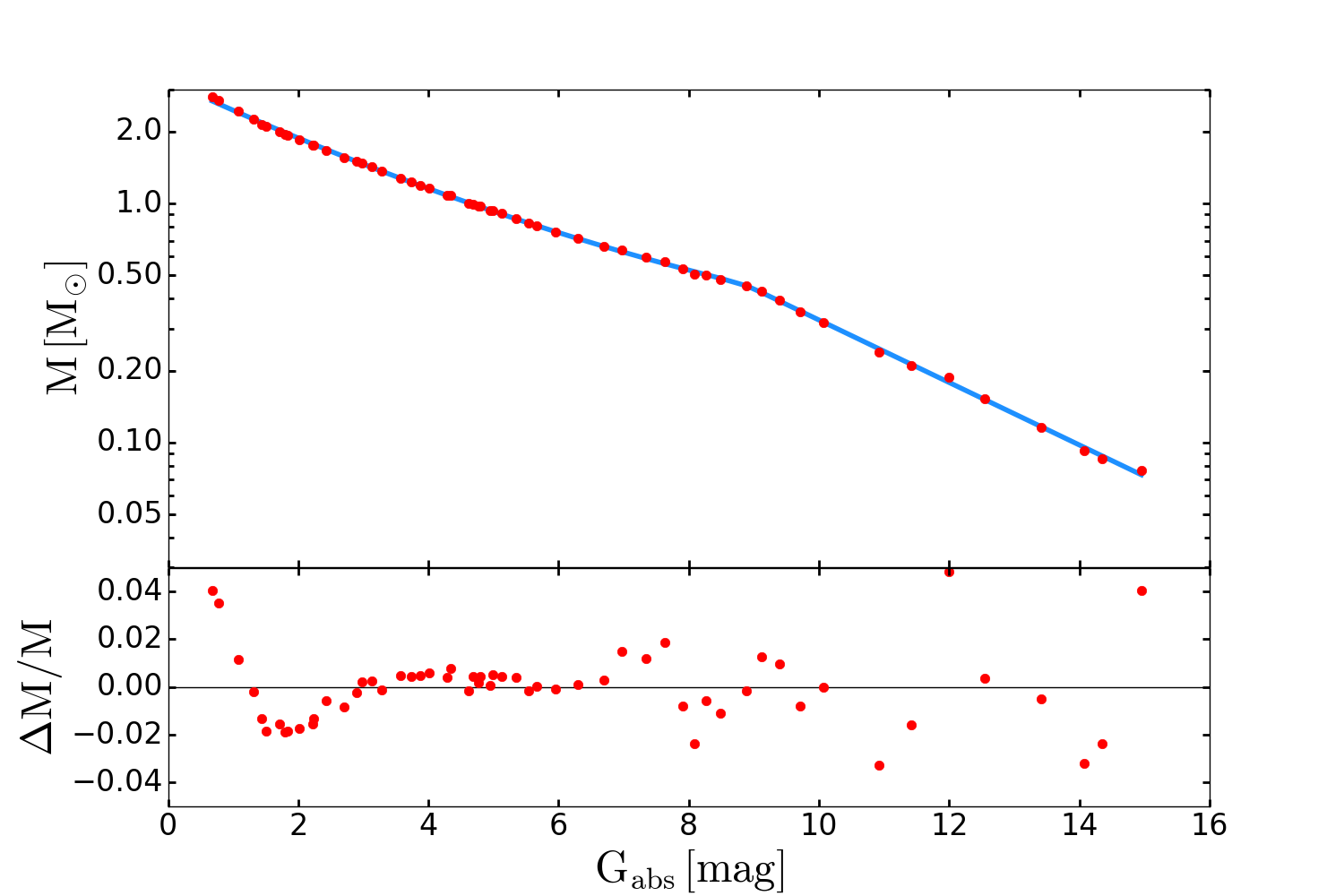}
\caption{Fitted \(G_{abs}\)- mass  relation. The red points show  the derived masses for different stellar types. The blue line shows the fitted relation. The two slopes are caused by the different luminosity mass relations.  
In the bottom part, the relative residuals after the fit are shown.}
\label{figure:fit}
\end{figure}

For the main-sequence stars, we determined a relation between  G magnitudes and stellar masses.
We started with a list of  temperatures, stellar radii, absolute V magnitudes, and V-Ic colours for different stellar types on the main sequence \citep{2013ApJS..208....9P}. We then translated these relations into the Gaia filter system using the  colour relation from  \cite{2010A&A...523A..48J},

\begin{equation}
G-V  = -0.0257 - 0.0924  (V-Ic)  -0.1623 (V-Ic)^{2} + 0.0090 (V-Ic)^{3}
.\end{equation} 
For the  different stellar types, we calculated the stellar masses using the  luminosity equation 
\begin{equation}
 \frac{L}{L_{\odot}} = \left(\frac{R}{R_{\odot}}\right)^{2} \left(\frac{T}{T_{\odot}}\right)^{4},
\end{equation} 
and the mass-luminosity relations  \citep{2005essp.book.....S}

\begin{equation}
\begin{aligned}
\text{for }L <  0.0304: &\,\frac{L}{L_{\odot}} = 0.23 \left(\frac{M}{M_{\odot}}\right)^{2.3},\\
\text{for }L > 0.0304: &\,\frac{L}{L_{\odot}} =  \left(\frac{M}{M_{\odot}}\right)^{4}        
\end{aligned}
.\end{equation} 
Finally, we fitted two exponential functions to the data and got the  equations 

\begin{equation}
\begin{aligned}
\text{for } G_{abs} < 8.85:\\ 
\log\left(\frac{M}{M_{\odot}}\right)   &= 0.00786\,G_{abs}^{2} -0.290\,G_{abs} + 1.18, \\
\text{for } 8.85 <  G_{abs} < 15:\\     
\log\left(\frac{M}{M_{\odot}}\right)& =         -0.301\,G_{abs}  + 1.89.
\end{aligned}
\label{Eq:Gmag} 
\end{equation} 
In Fig. \ref{figure:fit}  the fitted relation and its residuals are displayed. The relative residuals in the interesting regime (\(\sim 2< G_{abs} <\sim15\))  are below \(2\%\), which is amply sufficient for our purpose. However, in the error calculus below, we consider a mean error of \(10\%\) to account also for the uncertainties in  G magnitude, parallax, in the equations used and in the dependence on metallicity. 
We do not use a relation based on Gaia colours, for two reasons: first,  some of our lenses do not have colour information in DR2, and second, our sample contains many metal-poor halo stars. Hence they appear much bluer, whereas the change in absolute magnitude is small.
For  \(G_{abs} > 15.0\) ( i.e. \(M <  \sim 0.07 M_{\odot}\)) we reach the area of brown dwarfs. Those stars cannot be described by the mass-luminosity relation.  Hence, for them  we chose a fixed mass of \((0.07 \pm 0.03)\,M_{\odot}\).
 
We note that all of the calculated masses are only rough estimates in order to get an expectation of the Einstein radii,  astrometric shifts, and magnifications of the forecast microlensing events. An exact and direct determination of their masses would not be a pre-requisite, but the goal of observing these events.   

Using the estimated masses \(M_{L}\) and the {\it Gaia} DR2  parallaxes \(\varpi\) and \(\text{Sou\_}\varpi,\) we calculated the Einstein radii via the rewritten Eq. (\ref{equation:theta_E}),
\begin{equation}
\theta_{E} = \sqrt{\frac{4GM_{L}}{c^{2}}\frac{\varpi-\text{Sou\_}\varpi}{1pc \cdot 1''}} = 2.854\,mas \sqrt{\frac{M_{L}}{M_{\odot}}\cdot\frac{\varpi-\text{Sou\_}\varpi}{1\,mas}}
.\end{equation}
Finally, we computed the expected  shifts (\(\delta\theta_{c}\), \(\delta\theta_{+}\) and \(\delta\theta_{c,lum}\)) and  magnifications based on the equations in Sect. \ref{chapter:microlensing}.  We only selected those candidates where \(\delta\theta_{+} > 0.1\,\mathrm{mas}\).

\begin{table}[]
\caption{Our quality cuts applied to the raw target list of high-proper-motion stars, background sources, and events. These quantetites based on the position (\(ra,\,dec\), the total proper motion (\(\mu_{tot}\)), the parallax (\(\varpi\)), the number of photometric observations in G (\(n_{obs}\))  by Gaia, the G flux (\(G\_flux\)),  the corresponding errors (\(\sigma_{...}\)) as well as the expected shift of the brightest image (\(\delta\theta_{+}\)). Parameters  from the background source are indicated with a \(\text{Sou}\_\) prefix.} 
\label{tab:cuts}
\begin{tabular}{|l|r|}
\hline
application & criteria   \\
\hline
Lenses&\(\mu_{tot}>150\,\mathrm{mas/yr}\)\\
Lenses&\(\varpi/\sigma_{\varpi}>8\)\\
Lenses&\(\varpi/\mu_{tot} < 0.3\,\mathrm{yr}\)\\
Lenses&\(n_{obs}^{2} \cdot G\_flux / \sigma_{G\_flux} > 10^{6}\)\\
\hline
Sources&\((\text{Sou}\_\varpi+0.029\,\mathrm{mas})/\text{Sou}\_\sigma_{\varpi}> - 3\)\\
Sources&\(\sqrt{\text{Sou}\_\sigma_{ra}^2+\text{Sou}\_\sigma_{dec}^2}< 10\,\mathrm{mas}\)\\
 Sources&\(\lvert \vec{\mu}_{tot} - \text{Sou}\_\vec{\mu}_{tot}\rvert\,  < 0.7\cdot \lvert  \vec{\mu}_{tot} \rvert\) \\
Sources&\(\text{Sou}\_\varpi< \varpi \)\\

\hline
Events & \(\delta\theta_{+} > 0.1\,\mathrm{mas}\) \\

\hline

\end{tabular}
\end{table}

\begin{figure*}[]
\center
\includegraphics[width = 18 cm]{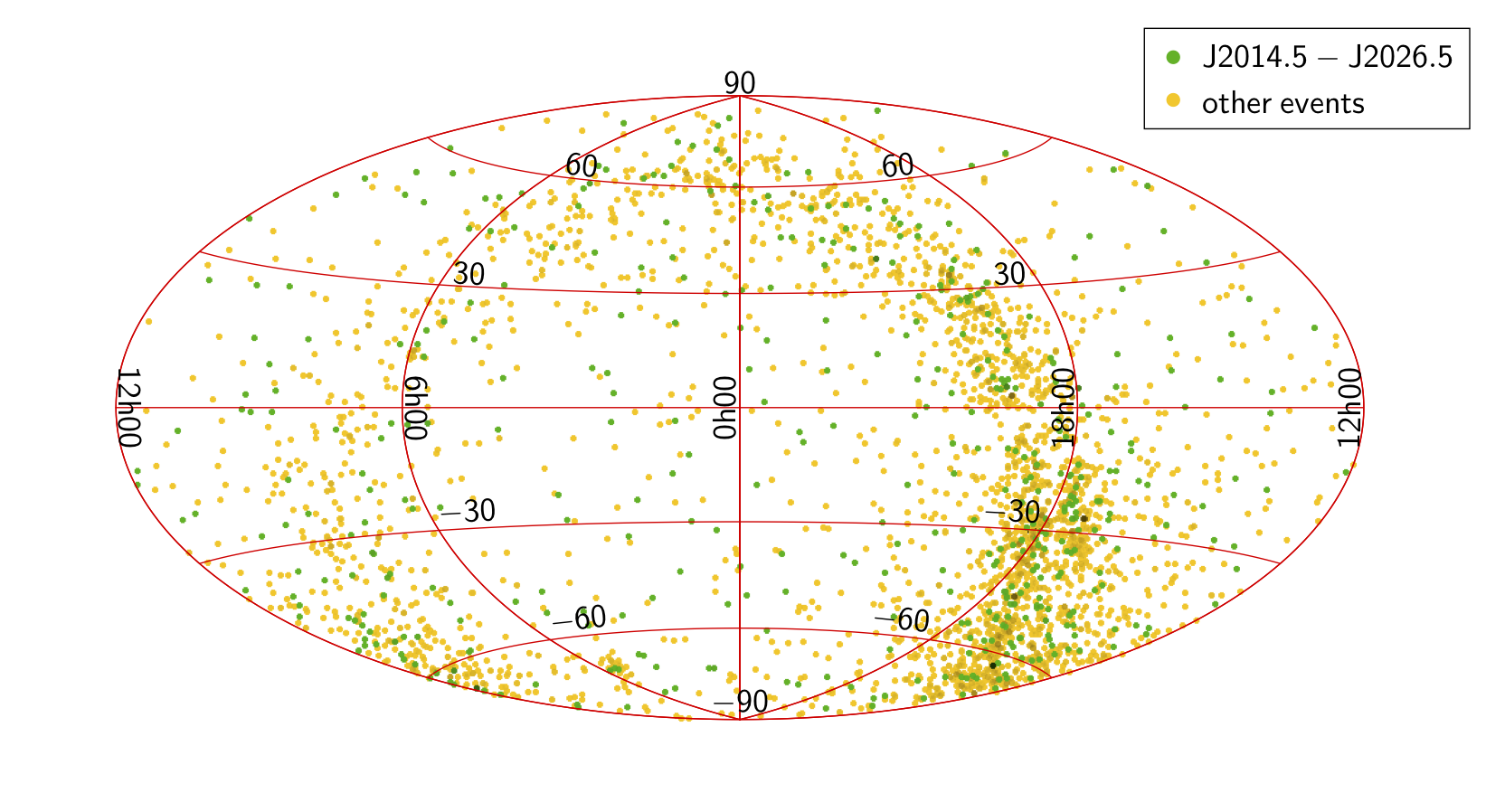}
\caption{Aitoff projection in equatorial coordinates of the events  between 2014.5 and 2026.5 (green) and all other events (yellow). Most of the events are in the galactic plane. }
\label{fig:Aitoff_events}
\end{figure*}

\section{Results: Astrometric microlensing events} 
\label{chapter:results}

We report the prediction of 3914 microlensing events by 2875 different lenses between J2010.0 and J2065.5. The past events are still of interest since  Gaia possibly measured the shift of  those events already. 
Due to the (small) motion of the background sources, some of them have a closest approach outside our original search interval in time (five earlier than J2010.0, 49 later than J2065.5)

The properties of our sample and a few interesting events are discussed  in the following. In Table \ref{tab:result},
 30 particularly interesting events are shown.  
The full catalogue of microlensing events can be accessed through the GAVO Data Center\footnote{German Astrophysical Virtual Observatory,\\ \url{http://dc.zah.uni-heidelberg.de/amlensing/q2/q/form}.}, and through Virtual Observatory (look for ``Astrometric Microlensing Events Predicted from Gaia DR2'').
In the following,  ``shift'' refers to the astrometric displacement of the brightest image only (\(\delta\theta_{+}\)) and  ``shift of the centre of light'' refers to the combined centre of light (\(\delta\theta_{c,lum}\)) considering the luminous-lens effect.

\subsection{The full sample}
Figure \ref{fig:Aitoff_events} shows the distribution of all our 3914 events on the sky.
Most of those are located towards the Galactic plane or the Large Magellanic Cloud, due to the high density of available background stars.  
For 1139 of the events the expected shift is smaller than three times its standard error.
Insignificant shifts are mainly caused by the uncertainties in the positions of the sources due to their unknown proper motion 
or smallest separations below \(100\,\mathrm{mas}\).

Inspecting the G magnitude difference, in 210 events the  source is  brighter than the lens. Among the rest, 726 events have a source less than three magnitudes
fainter than the lens, and for a total of 1050 events the sources are between three magnitudes and six magnitudes fainter than the lens. The remaining 1928 sources are more than six magnitudes fainter than the lens. These  magnitude differences  will change for different filters. The bright lenses tend to have large Einstein radii. Hence a measurable  shift is also expected at larger separations, where the source might  be detectable next to a bright star, even when the source is more than six magnitudes fainter. \cite{2018arXiv180701318Z} have shown that such observations are possible.

\begin{figure*}
\center
\includegraphics[width = 18. cm]{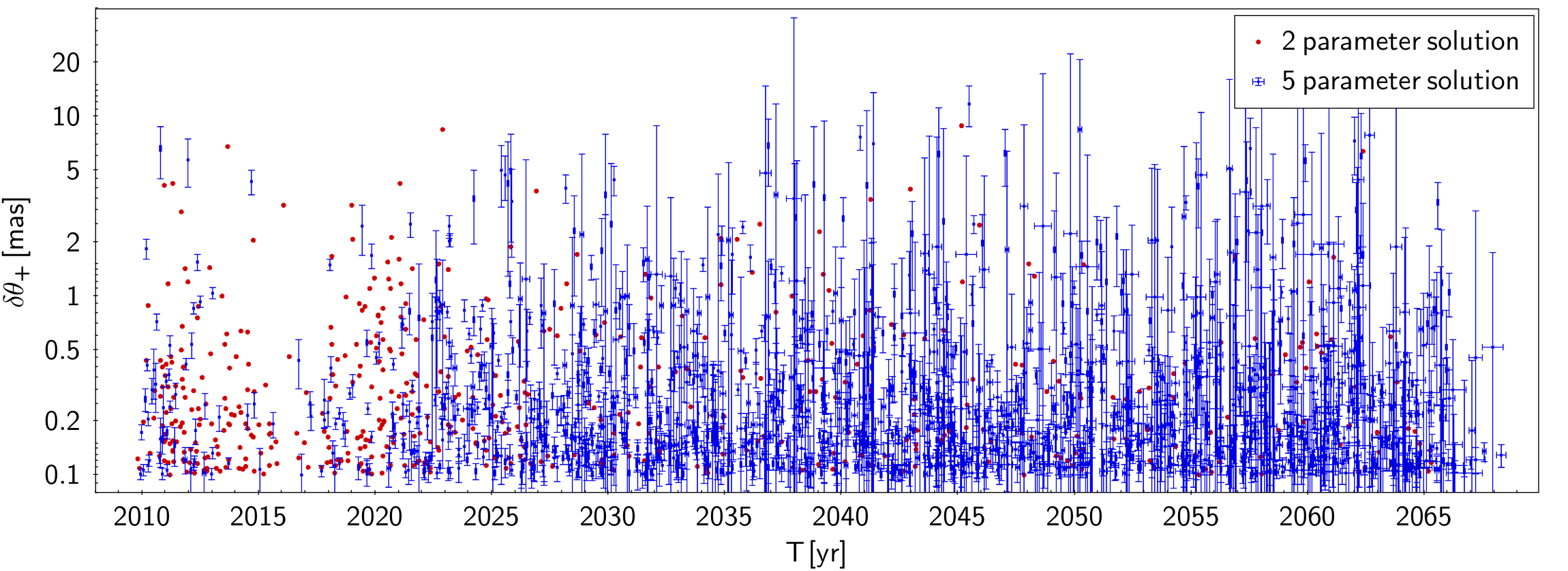}
\caption{Expected maximum shifts for all events with a source less than six magnitudes fainter than the lens. The red dots indicate the events where the proper motion and parallax of the source is unkown. The blue dots show the events with a five-parameter solution for the background sources, as well as the determined standard errors. The apparent paucity of events during the Gaia mission time is due to the angular resolution limit of {\it Gaia} DR2.}
\label{fig:shift_all}
\end{figure*}

The following numbers refer to the sample of 210 + 726 + 1050 events with a magnitude difference below \(6\,\mathrm{mag}\). 
Figure \ref{fig:shift_all} shows the date of the closest approach and the expected astrometric shift. 

For 431, 201, and 54 \textbf{events}, respectively,  we expect a shift of the brightest image larger than \(0.5\,\mathrm{mas}\), \(1\,\mathrm{mas,}\)  and \(3\,\mathrm{mas}\), respectively.  Of them, 88, 18 and two have a minimum separation larger than  \(100\,\mathrm{mas, respectively}\).
For 679 of the events the smallest separation is below \(100\,\mathrm{mas}\).  Considering luminous-lens effects, 198, 44, and  18 of those events have an expected shift of the centre of light larger than \(0.1\,\mathrm{mas}\), \(0.5\,\mathrm{mas}\)  and \(1\,\mathrm{mas}\), respectively.  We note that the luminosity effects depend on the used filters, and modern telescopes with adaptive optics or interferometry can even resolve  separations smaller than \(100\,\mathrm{mas}\). 

\cite{2018arXiv180610003B} predicted  2509 events until the year 2100. Due to different selection criteria and time ranges, we only detect  656 of their events independently. For all common events, the predicted dates and impact parameters are similar, within the standard errors. 

\subsection{Photometric microlensing effects of our astrometric microlensing events.}
\begin{figure}
\includegraphics[width = 9 cm]{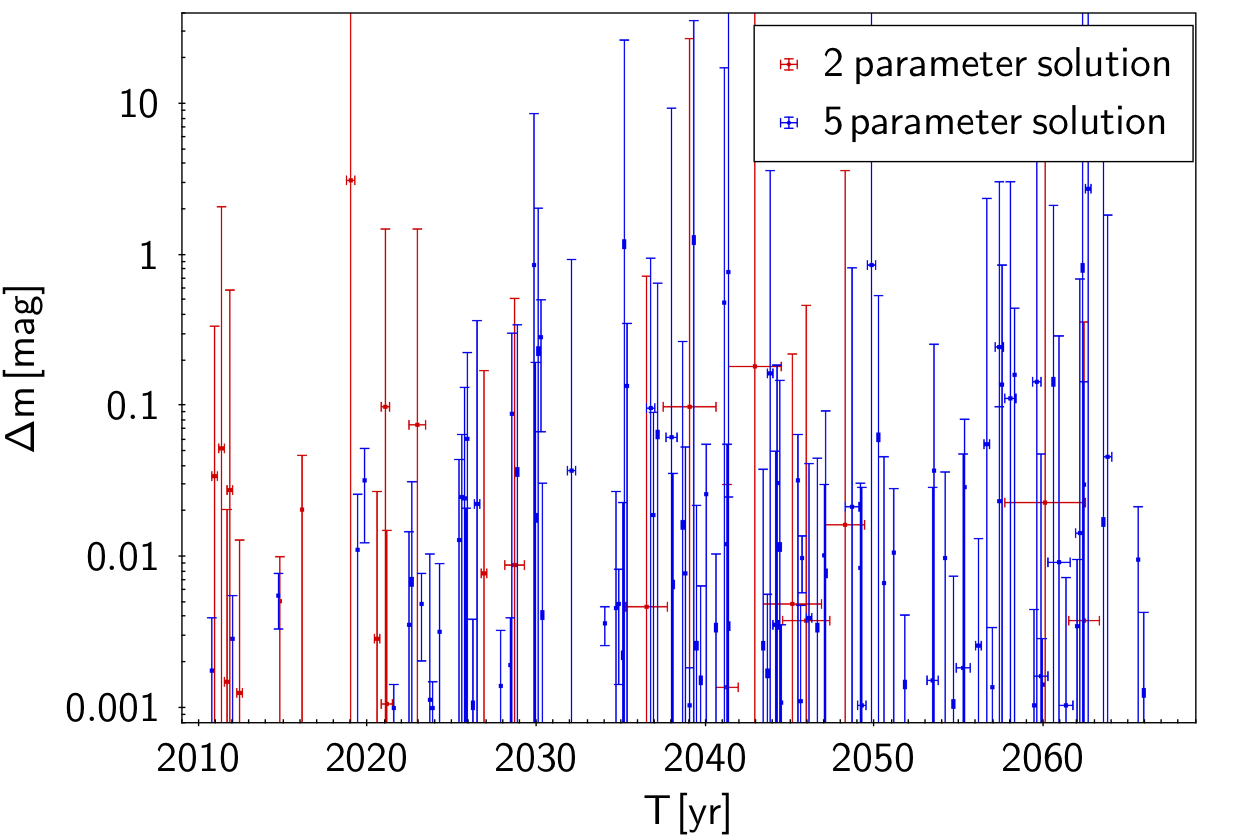}
\caption{Expected maximum magnification for all photometric events, with (blue) and without (red) five-parameter solution for the background source in {\it Gaia} DR2.}
\label{fig:mag}
\end{figure}

\cite{2018arXiv180511638M} recently reported 30 possible photometric microlensing events in the next 20 years  (J2015.5 to J2035.5).
Twenty-four of their candidates are also listed in our sample, the other six have an absolute proper motion below \(150\,\mathrm{mas/yr}\).  We found 246 events in the same time range  with a magnification greater than  \(A_{lum} -1 > 10^{-7}\) , which is the lowest magnification of their candidates.  
The typical photometric precision of photometric microlensing surveys is on the order of a few milli-magnitudes  \citep{2015AcA....65....1U}. Hence, we assume a limit of \(1\,\mathrm{mmag}\) to talk about photometric microlensing events.
This criterion is only fulfilled for five of their events.
For the same five events, a shift of the combined centre of light above \(0.1\,\mathrm{mas}\) is expected.
In our sample, 127 events fulfil this criterion, and for 20 events the magnification is above \(0.1\,\mathrm{mag}\).  For 104 and 18 of those, respectively, the motion of the background source is not known.

For all of our photometric events, Fig. \ref{fig:mag} shows the magnification and the predicted date. 
Since the predicted separation has to be really small, of the order \(\Theta_{E}\), in order to produce  a photometric effect, almost all predicted magnifications are not significant, especially when {\it Gaia} DR2 provides only a two-parameter solution for the background source. Furthermore, for many of these photometric events also the difference between the L2 magnification and the magnification seen from Earth is measurable.

\subsection{Candidates during the Gaia mission}
\begin{figure}
\includegraphics[width = 9 cm]{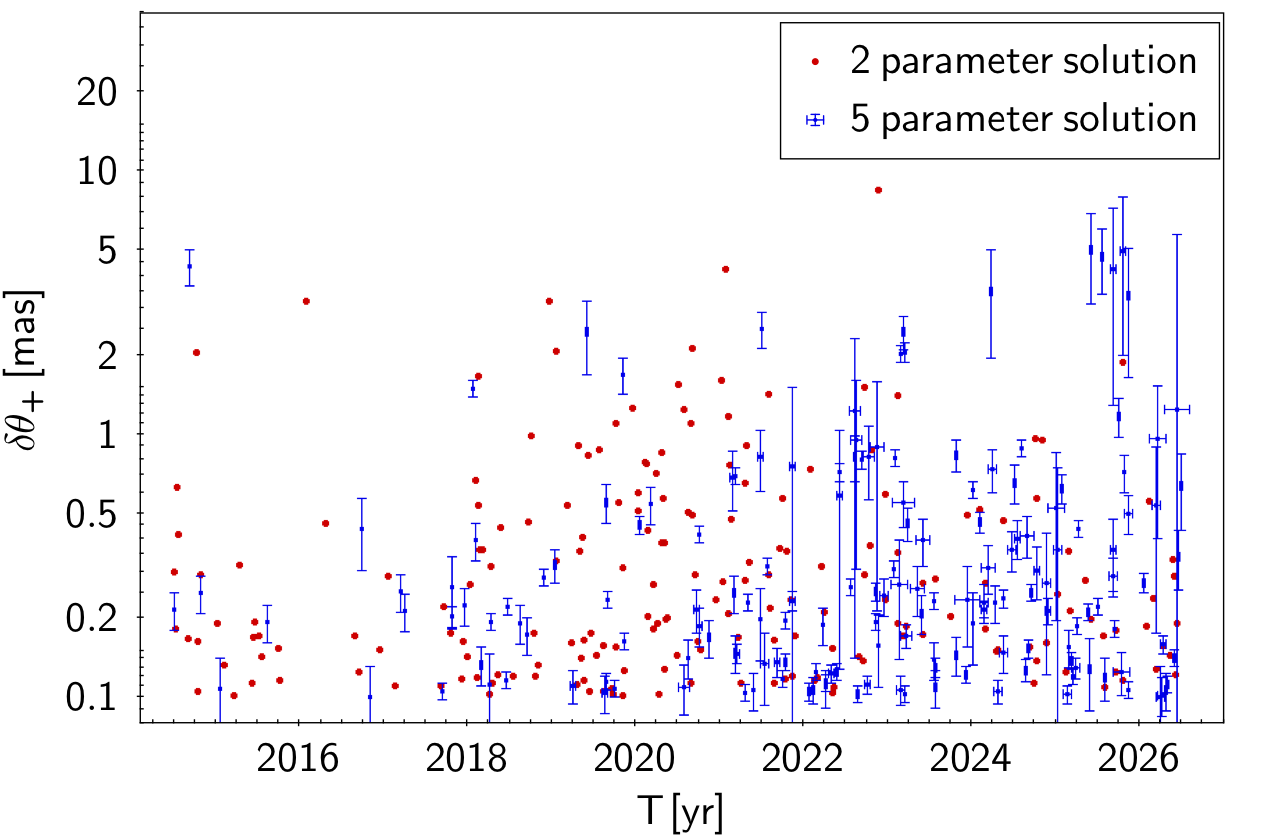}
\caption{Maximum shifts  for all expected  events between 2014.5 and 2026.5, with (blue) and without (red) five-parameter solution for the background source in {\it Gaia} DR2.}
\label{fig:shift_gaia}
\end{figure}
Since Gaia obtains many precise measurements over its mission time (from J2014.5 up to possibly J2024.5), events during this time are of special interest. During a slightly extended  period of time (2026.5; to accommodate events starting during the late Gaia mission), we found 544 events with an astrometric shift above \(0.1\,\mathrm{mas}\). For only 245 events,  proper motions and parallaxes of the sources are known. The numbers for those events will be given in parentheses in the following. 
 Of the events, 147 (62) have a minimum separation  below \(100\,\mathrm{mas}\) and will be (or were) blended for Gaia during the closest approach in the along-scan direction. In the across-scan direction they will be blended for a more extended time interval.
For 44 (19) events the shift of the blended centre of light is larger than \(0.1\,\mathrm{mas}\). 
For 29 (19) events we expect also a measurable  magnification above \(1\,\mathrm{mmag}\). 
The epoch and the astrometric shifts for our candidates during the Gaia mission are shown in Fig. \ref{fig:shift_gaia}. Since the expected timescales are on the order of a few years. it might be possible that Gaia observes the beginning or end of an event with a closest approach before 2014.5 or after 2024.5. 

\cite{2018arXiv180510630B} has recently reported 76 events during the Gaia mission life  time (between J2014.5 and J2026.5).
Independently, we discovered 60 of his events. The dates and  distances of the common events are similar except for ten events where \cite{2018arXiv180510630B} listed the dates close to J2026.5 or J2014.5 and we expect the date a few years later or earlier. 
For events where the proper motion of the background source is known, also the given uncertainties are similar. In the case of unknown  proper motions, our error estimates are much larger, since we assume an error of \(\text{Sou\_}\sigma_{\mu_{ra,dec}} = 10\,\mathrm{mas/yr}\).
The events which we did not reproduce either have a total proper motion of the lens below 150 mas/yr (eight cases), a positional uncertainty of the source above 10 mas (three cases), comparable proper motions (three cases), or are outside our defined box (two cases),   we deliberately excluded from our sample.

\subsection{White dwarfs}
Our catalogue contains 486 events caused by 352 different white dwarfs. For 427 of those, the background source is less than six magnitudes fainter in the G band. Since white dwarfs are blue objects, using infrared filters will be more advantageous for possible follow-up observations. Of the events, 84 will happen between 2014.5 and 2026.5. 
For 98 of the events, the expected maximum shift is above \(0.5\,\mathrm{mas}\) and for 53 above \(1\,\mathrm{mas}\)  (17 and 5 for the period 2014.5-2026.5).
For 22 events also the blended centre of light will be shifted by at least \(0.5\,\mathrm{mas}\).
We also independently recovered the events of WD 1142-645 predicted by  \cite{2018MNRAS.478L..29M}, and the one of Stein 51B, which was already observed by \cite{2017Sci...356.1046S}.

\subsection{Proxima Centauri  - the nearest }
\cite{2014ApJ...782...89S} predicted two microlensing events of Proxima Centauri in October 2014 and February 2016 with a closest separation of \(1600\,\mathrm{mas}\) and \(500\,\mathrm{mas}\), respectively. By observing those events with VLT/SPHERE\footnote{Very Large Telescope equipped with the SPHERE instrument}  and HST/WFC3\footnote{Hubble Space Telescope equipped with the Wide Field Camera 3}, \cite{2018arXiv180701318Z} were able to determine the mass of Proxima Centauri, but with an uncertainty of about 40\%.
We did not recover either of those two events, since the background stars are not listed in {\it Gaia} DR2. However, we found 84 further microlensing events of Proxima Centauri until J2065.5. 
Nine of those have an expected shift larger than \(1\,\mathrm{mas}\).  
With a G magnitude of \(8.9\,\mathrm{mag,}\)  Proxima Centauri is much brighter than the sources \((\Delta m  \sim 6 - 12\,\mathrm{mag})\). Therefore a significant shift of the centre of light of the blended system cannot be observed. Due to the large Einstein radius of Proxima Centauri (\(\Theta_{E} =\sim 27.1 \mathrm{mas}\)), a shift of \(1\,\mathrm{mas}\) can still be observed at a separation of \(\sim 700\,\mathrm{mas}\)  and for  all sources with a separation smaller than \(7000\,\mathrm{mas}\)  a shift larger than \(0.1\,\mathrm{mas}\) is expected. At this separation it is possible to observe  background stars next to Proxima Centauri.
 
\subsection{Barnard's star - the fastest}
Barnard's star is the fastest star on the sky. Hence the sky area passed by this star is the largest in our sample. Between J2010.0 and J2065.5  we found 37 astrometric microlensing events for Barnard's star.  Seven of those happen between 2014.5 and 2026.5, and so  Gaia might measure the deflections.
Barnard's star has a  G magnitude of 8.2 mag, and we determined an Einstein radius of \(\sim 28.6\,\mathrm{mas}\).
Due to its brightness, most of the sources are  more than six magnitudes fainter. However, in 2035 it will pass by a \(G = 11.8\,\mathrm{mag}\) star with a closest separation of \((335 \pm 13) \,\mathrm{mas}\). If it is possible to resolve source and lens, a shift of  \((2.43  \pm 0.20)\,\mathrm{mas}\) is expected. The shift of the blended centre of light will be smaller than \(0.1 \,\mathrm{mas}\). More information is given  in Table \ref{tab:result}, event 10.

\subsection{Two photometric events in 2019}
 In June 2019, a \(G = 15.2\,\mathrm{mag}\) star  ({\it Gaia} DR2 source id: 5862333044226605056) will pass a \(G = 18.1\,\mathrm{mag}\) star with a closest separation of \((6.48\pm 3.4)\,\mathrm{mas}\). For this event we determined an Einstein radius of \((4.66\pm 0.24)\,\mathrm{mas}\).
The blended centre of light will be shifted by \((0.18\pm 0.06)\,\mathrm{mas}\), and we expect a magnification of \((0.011\pm 0.014)\,\mathrm{mag}\).  In November 2019 we expect a second photometric event when the \(G = 17.2 \,\mathrm{mag}\) star 
2MASS J13055171-7218081 ({\it Gaia} DR2 source id: 5840411363658156032) passes  a \(G = 18.2 \,\mathrm{mag}\) star with a closest separation of \((5.83 \pm 1.32)\,\mathrm{mas}\). We determined an Einstein radius of \((3.56\pm0.18)\,\mathrm{mas}\).  The event will be magnified by \((0.032 \pm 0.019)\,\mathrm{mag}\), and the expected shift of the combined centre of light will be \((0.50 \pm 0.08)\,\mathrm{mas}\). Both events are listed in Table \ref{tab:result} (lines 3 and 4).
The event of 2MASS J13055171-7218081 was also predicted by \cite{2018arXiv180510630B} independently.

\subsection{Two astrometric  events in 2018}
In  \cite{2018arXiv180508023K}, 
we already reported two ongoing astrometric microlensing events  in Summer  2018 by Luyten 141-23 and Ross 322

%

\section{Conclusion}
 \label{chapter:conclusion}

We determined a list of 148 000 high-proper-motion stars using {\it Gaia} DR2. We then searched for background sources close to their paths and found \(\sim 68000\) candidates for astrometric microlensing events.
For those, we computed the closest projected distances and the expected astrometric and photometric effects.
The main difficulty in this process is to sort out probably erroneous DR2 data, while losing as few as possible valid events. We chose the rejection criteria such as to be confident that our list shows a small false positive rate, while not deleting too many promising predictions.
  
Because of the large sample, we were not able to perform a Monte Carlo simulation to determine the uncertainties of our predictions, due to limited resources. Instead, we used an error propagation, which leads to robust values as long as the relative errors are small. At small separations, the derived uncertainties for shift and magnification tend to be overestimated.  

In total, we give predictions for 3914 microlensing events caused by 2875 different lens stars with an expected shift of the brighter image larger than \(0.1\,\mathrm{mas}\). These include about 700 events, which were also predicted by \cite{2018arXiv180510630B}, \cite{2018arXiv180511638M}, and \cite {2018arXiv180610003B}. 
The independent detection of those shows the reliability of  the respective methods.

The standard errors of the predicted  date of the closest approach is a few weeks for most of the events 
(and for the best events only a few hours). 
This is much smaller than the duration of the events.
The standard error for the minimum separation  is typically on the order of a few dozen milliarcseconds.
Large uncertainties are mostly caused by unknown proper motions of the source.
As expected, the standard errors  increase with the time before and after J2015.5. 
However, for the year 2065, it is still possible to predict events with a separation significantly below \(100 \mathrm{mas}\) (\(d+3\sigma_{d} < 100\,\mathrm{mas}\)). 

Typically the lens is much brighter than the source for 1928 events even with a G magnitude difference above 6 mag. These are hard to observe, but using more suitable wavelength bands the  brightness differences can be reduced.

Observations of the events and the  determination of the lens masses will lead to a better understanding of mass relations for main sequence stars  \citep{1991ApJ...371L..63P}. 
Perhaps even more interestingly, our sample contains also events caused by 352 different white dwarfs. The observation and subsequent mass determination  of those will lead to a better understanding of white dwarfs, and the final phase of the evolution of stars.

With {\it Gaia} DR 3 (expected in late 2020) we expect an improvement of the standard errors. In other words, the precision of the predicted events will increase.
In addition, the number of background sources with five-parameter solutions will increase, which again leads to better predictions.
Furthermore, {\it Gaia} DR3 will include  detections and treatment of binary stars, while for {\it Gaia} DR2 all stars were treated as single. This too will help to make even more precise predictions for the paths of the lenses. 
With {\it Gaia} DR4 (expected  in late 2022) the individual astrometric Gaia measurements will be published. 
Using such data it will be  possible to perform a detailed reconstruction and modeling of past events and to even more precisely determine the masses of their lenses.

\begin{acknowledgements}
This work has made use of results from the ESA space mission {\it  Gaia}, the data from which were processed by the {\it Gaia} Data Processing and Analysis Consortium (DPAC). Funding for the DPAC has been provided by national institutions, in particular the institutions participating in the {\it Gaia} Multilateral Agreement. The {\it  Gaia} mission website is:
http://www.cosmos.esa.int/Gaia. Some of the authors are members of the {\it  Gaia} Data Processing and Analysis Consortium (DPAC). This research has made use of the SIMBAD database,
operated at CDS, Strasbourg, France. This research made use of Astropy, a community-developed core Python package for Astronomy \citep{refId0}.
This research made use of  TOPCAT \citep{2005ASPC..347...29T}, which was used to prepare ten figures in this paper. 
We gratefully acknowledge the technical support we received from
the staff of the e-inf-astro project (BMBF F\"{o}rderkennzeichen 05A17VH2).

\end{acknowledgements}

\begin{sidewaystable*} 
\caption[]{List of 30 especially promising astrometric microlensing events. The table lists the Gaia Source ID of lens and background star,  the stellar type (\(ST\)) of the lens, its mass (\(M\)), the Einstein radius \(\theta_{E}\), the Julian year T after 2000 of the closest approach, with uncertainty, the minimum separation (\(d_{min}\)), as well as the expected shift of the combined centre of light (\(\delta\theta_{c,\, lum }\)), the shift of the brightest image only (\(\delta\theta_{+}\)), and  the expected magnification (\(\Delta m\)). The corresponding errors are indicated by \(\sigma_{...}\). The online table\footnote{\url{http://dc.zah.uni-heidelberg.de/amlensing/q2/q/form}.} will also list  positions, proper motions, parallaxes, and magnitudes for the source and lens stars, the timescale, the impact parameter in units of Einstein radii, the expected shift of the centre of light assuming a dark lens, as well as the predicted values as seen from L2. For the background source of event \#2, only a two-parameter solution exists. Events \#3 and \#4 (light grey) are the photometric events in 2019.  Event \#10 (dark grey) is the event of Barnard's star in 2035.}
\label{tab:result}
\small 
\begin{tabular}{l|rrrrrrrrrrrrrrrr|}
\#&\(source\,ID\)& \(\text{Sou\_}source\,ID\) & ST& \(M\)& 
\(\theta_{E}\) &\(\sigma_{\theta_{E}}\) &
\(T-2000\)&\(\sigma_{T}\)&
\(d_{min}\)&\(\sigma_{d_{min}}\)&
\(\delta\theta_{c,\,lum}\) & \(\sigma_{\delta\theta_{c,\,lum}}\) &
\(\delta\theta_{+}\) & \(\sigma_{\delta\theta_{+}}\)&
\(\Delta m\) & \(\sigma_{\Delta m }\) \\ 

& & &  & \(M_{\odot}\) & 
mas &mas &
Jyear &Jyear&
mas&mas&
mas&mas&
mas & mas&
mag & mag \\
\hline 
\(1 \) & \(478978296199510912 \) & \(478978296204261248 \) & \(WD       \) & \(0.65        \) & \(15.5  \) & \(1.8         \) & \(14.6923 \) & \(0.0024    \) & \(50.7 \) & \(1.1       \) & \(1.78     \) & \(0.27           \) & \(4.34      \) & \(0.67            \) & \(0.0055       \) & \(0.0023\)             \\
\(2 \) & \(4733794485572154752\) & \(4733794485572154624\) & \(BD       \) & \(0.07        \) & \(8.9   \) & \(1.9         \) & \(16.0744 \) & \(0.0099    \) & \(21.1 \) & \(7.7       \) & \(1.68     \) & \(0.61           \) & \(3.2       \) & \(1.3             \) & \(0.021        \) & \(0.027\)              \\
\rowcolor{lightGray}
\(3 \) & \(5862333044226605056\) & \(5862333048529855360\) & \(MS       \) & \(0.400      \) & \(4.66  \) & \(0.24        \) & \(19.417  \) & \(0.012     \) & 
\(6.4  \) & \(3.5       \) & \(0.184    \) & \(0.069          \) & \(2.44      \) & \(0.76            \) & \(0.011        \) & \(0.015\)              \\
\rowcolor{lightGray}
\(4 \) & \(5840411363658156032\) & \(5840411359350016128\) & \(MS       \) & \(0.174      \) & \(3.56  \) & \(0.18        \) & \(19.8392 \) & \(0.0029    \) & \(5.8  \) & \(1.3       \) & \(0.505    \) & \(0.079          \) & \(1.69      \) & \(0.27            \) & \(0.032        \) & \(0.020\)              \\
\(5 \) & \(4687445500635789184\) & \(4687445599404851456\) & \(WD       \) & \(0.65        \) & \(13.6  \) & \(1.7         \) & \(21.5001 \) & \(0.0062    \) & \(70.3 \) & \(1.9       \) & \(0.98     \) & \(0.16           \) & \(2.52      \) & \(0.41            \) & \(0.00101      \) & \(0.00044\)            \\
\hline
\(6 \) & \(4248799013208327424\) & \(4248799013215266176\) & \(MS       \) & \(0.219      \) & \(4.10  \) & \(0.23        \) & \(29.876  \) & \(0.039     \) & \(0.9  \) & \(10.       \) & \(0.9      \) & \(3.8            \) & \(3.7       \) & \(4.4             \) & \(0.9          \) & \(7.8\)                \\
\(7 \) & \(5918299904067162240\) & \(5918299908365843840\) & \(MS       \) & \(0.113      \) & \(6.93  \) & \(0.35        \) & \(30.2480 \) & \(0.0032    \) & \(6.3  \) & \(2.8       \) & \(2.17     \) & \(0.17           \) & \(4.46      \) & \(0.84            \) & \(0.28         \) & \(0.22\)               \\
\(8 \) & \(6130500670360253312\) & \(6130500567281038080\) & \(MS       \) & \(0.318       \) & \(2.16  \) & \(0.33        \) & \(35.162  \) & \(0.072     \) & \(0.2  \) & \(7.4       \) & \(0.4      \) & \(4.3            \) & \(2.0       \) & \(3.5             \) & \(1.           \) & \(24.\)                \\
\(9 \) & \(5863711561290571008\) & \(5863711561290570112\) & \(MS       \) & \(0.199      \) & \(3.39  \) & \(0.17        \) & \(35.333  \) & \(0.014     \) & \(5.0  \) & \(3.4       \) & \(1.179    \) & \(0.087          \) & \(1.70      \) & \(0.69            \) & \(0.13         \) & \(0.22\)               \\
\rowcolor{Gray} 
\(10\) & \(4472832130942575872\) & \(4472836292758713216\) & \(MS       \) & \(0.184      \) & \(28.7  \) & \(1.5         \) & \(35.76441\) & \(0.00073   \) & \(335. \) & \(13.       \) & \(0.0802   \) & \(0.0065         \) & \(2.43      \) & \(0.20            \) & 3.69E-6      & 9.3E-7             \\
\hline
\(11\) & \(6074397471079635328\) & \(6074397471080379776\) & \(WD       \) & \(0.65        \) & \(8.3   \) & \(1.1         \) & \(36.72   \) & \(0.26      \) & \(9.5  \) & \(40.       \) & \(1.7      \) & \(2.6            \) & \(5.        \) & \(10.             \) & \(0.10         \) & \(0.87\)               \\
\(12\) & \(5556349476589422336\) & \(5556349472293892224\) & \(WD       \) & \(0.65        \) & \(7.94  \) & \(0.96        \) & \(37.173  \) & \(0.077     \) & \(15.  \) & \(47.       \) & \(2.1      \) & \(3.2            \) & \(3.7       \) & \(8.2             \) & \(0.06         \) & \(0.57 \)               \\
\(13\) & \(5886942661427781760\) & \(5886942661374132096\) & \(MS       \) & \(0.180      \) & \(3.69  \) & \(0.19        \) & \(39.266  \) & \(0.049     \) & \(0.2  \) & \(13.       \) & \(0.5      \) & \(8.0            \) & \(3.5       \) & \(5.9             \) & \(1.           \) & \(35.\)                \\
\(14\) & \(5715906236031073280\) & \(5715906236031079040\) & \(WD       \) & \(0.65        \) & \(7.85  \) & \(0.91        \) & \(40.049  \) & \(0.034     \) & \(20.0 \) & \(6.9       \) & \(2.06     \) & \(0.51           \) & \(2.71      \) & \(0.84            \) & \(0.026        \) & \(0.030\)              \\
\(15\) & \(5332606522595645952\) & \(5332606277747043456\) & \(WD       \) & \(0.65         \) & \(33.7  \) & \(3.9         \) & \(40.7896 \) & \(0.0021    \) & \(139.6\) & \(4.9       \) & \(0.0470   \) & \(0.0077         \) & \(7.7       \) & \(1.3             \) & 3.4E-5        & 1.6E-5            \\
\(16\) & \(4118914220102650624\) & \(4118914185707335040\) & \(MS       \) & \(0.319       \) & \(7.11  \) & \(0.36        \) & \(41.383  \) & \(0.014     \) & \(0.1  \) & \(13.       \) & \(0.07     \) & \(5.1            \) & \(7.1       \) & \(6.5             \) & \(0.8          \) & \(92.\)                \\
\(17\) & \(428051391503714432 \) & \(428051391509474816 \) & \(WD       \) & \(0.65        \) & \(8.18  \) & \(0.97        \) & \(43.61   \) & \(0.11      \) & \(47.  \) & \(29.       \) & \(1.27     \) & \(0.71           \) & \(1.39      \) & \(0.83            \) & \(0.0017       \) & \(0.0040\)             \\
\(18\) & \(5605383430285597696\) & \(5605383537671689856\) & \(WD       \) & \(0.65        \) & \(15.0  \) & \(1.7         \) & \(43.9735 \) & \(0.0091    \) & \(207.6\) & \(4.5       \) & \(1.04     \) & \(0.17           \) & \(1.08      \) & \(0.18            \) & 5.6E-5        & 2.7E-5             \\
\(19\) & \(6282457918962299776\) & \(6282457815883084928\) & \(WD       \) & \(0.65        \) & \(19.9  \) & \(2.3         \) & \(45.466  \) & \(0.012     \) & \(22.  \) & \(10.       \) & \(1.53     \) & \(0.43           \) & \(11.7      \) & \(3.0             \) & \(0.032        \) & \(0.033\)              \\
\(20\) & \(3365063724883180288\) & \(3365062964671171712\) & \(MS       \) & \(0.119       \) & \(11.08 \) & \(0.56        \) & \(45.607  \) & \(0.011     \) & \(119. \) & \(14.       \) & \(0.0095   \) & \(0.0013         \) & \(1.02      \) & \(0.14            \) & 1.46E-6  & 7.1E-7             \\
\hline
\(21\) & \(1822711900548572544\) & \(1822711900548570624\) & \(MS       \) & \(0.212       \) & \(2.37  \) & \(0.15        \) & \(49.80   \) & \(0.27      \) & \(0.3  \) & \(44.       \) & \(0.3      \) & \(17.            \) & \(2.        \) & \(20.             \) & \(0.8          \) & \(100. \)              \\
\(22\) & \(4203875751318123904\) & \(4203875648238893696\) & \(BD       \) & \(0.07        \) & \(5.1   \) & \(1.1         \) & \(53.538  \) & \(0.018     \) & \(11.  \) & \(22.       \) & \(1.2      \) & \(1.5            \) & \(2.1       \) & \(3.1             \) & \(0.04         \) & \(0.22\)               \\
\(23\) & \(4117081643422165120\) & \(4117081467277401344\) & \(WD       \) & \(0.65        \) & \(11.7  \) & \(1.4         \) & \(56.982  \) & \(0.027     \) & \(66.  \) & \(25.       \) & \(1.45     \) & \(0.54           \) & \(1.99      \) & \(0.77            \) & \(0.0014       \) & \(0.0021\)             \\
\(24\) & \(6126095232211644160\) & \(6126095300931121024\) & \(WD       \) & \(0.65        \) & \(5.68  \) & \(0.73        \) & \(57.33   \) & \(0.27      \) & \(4.6  \) & \(32.       \) & \(1.52     \) & \(0.22           \) & \(3.8       \) & \(9.8             \) & \(0.2          \) & \(2.9 \)               \\
\(25\) & \(6426625402561169024\) & \(6426625402559392896\) & \(MS       \) & \(0.272      \) & \(4.94  \) & \(0.25        \) & \(58.252  \) & \(0.017     \) & \(4.5  \) & \(4.2       \) & \(1.12     \) & \(0.20           \) & \(3.2       \) & \(1.3             \) & \(0.16         \) & \(0.28\)               \\
\hline
\(26\) & \(5243594081269535872\) & \(5243594253068231168\) & \(MS       \) & \(0.175      \) & \(14.77 \) & \(0.74        \) & \(58.8661 \) & \(0.0076    \) & \(208.4\) & \(5.8       \) & \(0.00767  \) & \(0.00059        \) & \(1.042     \) & \(0.079           \) & 3.93E-7 & 8.9E-8             \\
\(27\) & \(4484348145137238016\) & \(4484348145137243904\) & \(BD       \) & \(0.07        \) & \(3.39  \) & \(0.72        \) & \(60.595  \) & \(0.097     \) & \(4.   \) & \(28.       \) & \(1.06     \) & \(0.96           \) & \(1.8       \) & \(6.3             \) & \(0.1          \) & \(2.1\)                \\
\(28\) & \(6082407619449932032\) & \(6082407619449930880\) & \(MS       \) & \(0.560      \) & \(6.04  \) & \(0.31        \) & \(62.295  \) & \(0.050     \) & \(0.07 \) & \(8.8       \) & \(0.07     \) & \(3.8            \) & \(6.0       \) & \(4.4             \) & \(0.8          \) & \(69.\)                \\
\(29\) & \(2025071788687899776\) & \(2025071792959778688\) & \(WD       \) & \(0.65        \) & \(8.06  \) & \(0.94        \) & \(62.64   \) & \(0.20      \) & \(0.3  \) & \(30.       \) & \(2.       \) & \(27.            \) & \(8.        \) & \(14.             \) & \(2.           \) & \(101.\)               \\
\(30\) & \(4293315765165489536\) & \(4293315662070462080\) & \(BD       \) & \(0.07        \) & \(9.8   \) & \(2.1         \) & \(65.727  \) & \(0.051     \) & \(79.  \) & \(76.       \) & \(0.0059   \) & \(0.0059         \) & \(1.19      \) & \(1.2             \) & 2.3E-6        & 8.8E-6             \\
   \end{tabular} 
\normalsize
\end{sidewaystable*}



\begin{thebibliography}{40}
\expandafter\ifx\csname natexlab\endcsname\relax\def\natexlab#1{#1}\fi

\bibitem[{{Bond} {et~al.}(2001){Bond}, {Abe}, {Dodd}, {Hearnshaw}, {Honda},
  {Jugaku}, {Kilmartin}, {Marles}, {Masuda}, {Matsubara}, {Muraki}, {Nakamura},
  {Nankivell}, {Noda}, {Noguchi}, {Ohnishi}, {Rattenbury}, {Reid}, {Saito},
  {Sato}, {Sekiguchi}, {Skuljan}, {Sullivan}, {Sumi}, {Takeuti}, {Watase},
  {Wilkinson}, {Yamada}, {Yanagisawa}, \& {Yock}}]{2001MNRAS.327..868B}
{Bond}, I.~A., {Abe}, F., {Dodd}, R.~J., {et~al.} 2001, \mnras, 327, 868

\bibitem[{{Bramich}(2018)}]{2018arXiv180510630B}
{Bramich}, D.~M. 2018, ArXiv e-prints [\eprint[arXiv]{1805.10630}]

\bibitem[{{Bramich} \& {Nielsen}(2018)}]{2018arXiv180610003B}
{Bramich}, D.~M. \& {Nielsen}, M.~B. 2018, ArXiv e-prints
  [\eprint[arXiv]{1806.10003}]

\bibitem[{{Chwolson}(1924)}]{1924AN....221..329C}
{Chwolson}, O. 1924, Astronomische Nachrichten, 221, 329

\bibitem[{{Dominik} \& {Sahu}(2000)}]{2000ApJ...534..213D}
{Dominik}, M. \& {Sahu}, K.~C. 2000, \apj, 534, 213

\bibitem[{{Einstein}(1915)}]{1915SPAW...47..831E}
{Einstein}, A. 1915, Sitzungsber.~preuss.Akad.~Wiss., vol.~47, No.2,
  pp.~831-839, 1915, 47, 831

\bibitem[{{Einstein}(1936)}]{1936Sci....84..506E}
{Einstein}, A. 1936, Science, 84, 506

\bibitem[{{Fabricius} {et~al.}(2016){Fabricius}, {Bastian}, {Portell},
  {Casta{\~n}eda}, {Davidson}, {Hambly}, {Clotet}, {Biermann}, {Mora},
  {Busonero}, {Riva}, {Brown}, {Smart}, {Lammers}, {Torra}, {Drimmel},
  {Gracia}, {L{\"o}ffler}, {Spagna}, {Lindegren}, {Klioner}, {Andrei}, {Bach},
  {Bramante}, {Br{\"u}semeister}, {Busso}, {Carrasco}, {Gai}, {Garralda},
  {Gonz{\'a}lez-Vidal}, {Guerra}, {Hauser}, {Jordan}, {Jordi}, {Lenhardt},
  {Mignard}, {Messineo}, {Mulone}, {Serraller}, {Stampa}, {Tanga}, {van
  Elteren}, {van Reeven}, {Voss}, {Abbas}, {Allasia}, {Altmann}, {Anton},
  {Barache}, {Becciani}, {Berthier}, {Bianchi}, {Bombrun}, {Bouquillon},
  {Bourda}, {Bucciarelli}, {Butkevich}, {Buzzi}, {Cancelliere}, {Carlucci},
  {Charlot}, {Collins}, {Comoretto}, {Cross}, {Crosta}, {de Felice}, {Fienga},
  {Figueras}, {Fraile}, {Geyer}, {Hernandez}, {Hobbs}, {Hofmann}, {Liao},
  {Licata}, {Martino}, {McMillan}, {Michalik}, {Morbidelli}, {Parsons},
  {Pecoraro}, {Ramos-Lerate}, {Sarasso}, {Siddiqui}, {Steele},
  {Steidelm{\"u}ller}, {Taris}, {Vecchiato}, {Abreu}, {Anglada}, {Boudreault},
  {Cropper}, {Holl}, {Cheek}, {Crowley}, {Fleitas}, {Hutton}, {Osinde},
  {Rowell}, {Salguero}, {Utrilla}, {Blagorodnova}, {Soffel}, {Osorio},
  {Vicente}, {Cambras}, \& {Bernstein}}]{2016A&A...595A...3F}
{Fabricius}, C., {Bastian}, U., {Portell}, J., {et~al.} 2016, \aap, 595, A3

\bibitem[{{Gaia Collaboration} {et~al.}(2018){Gaia Collaboration}, {Brown},
  {Vallenari}, {Prusti}, {de Bruijne}, {Babusiaux}, \&
  {Bailer-Jones}}]{2018arXiv180409365G}
{Gaia Collaboration}, {Brown}, A.~G.~A., {Vallenari}, A., {et~al.} 2018, ArXiv
  e-prints [\eprint[arXiv]{1804.09365}]

\bibitem[{{Gaia Collaboration} {et~al.}(2016){Gaia Collaboration}, {Prusti},
  {de Bruijne}, {Brown}, {Vallenari}, {Babusiaux}, {Bailer-Jones}, {Bastian},
  {Biermann}, {Evans}, \& et~al.}]{2016A&A...595A...1G}
{Gaia Collaboration}, {Prusti}, T., {de Bruijne}, J.~H.~J., {et~al.} 2016,
  \aap, 595, A1

\bibitem[{{Gaudi}(2012)}]{2012ARA&A..50..411G}
{Gaudi}, B.~S. 2012, \araa, 50, 411

\bibitem[{{Hog} {et~al.}(1995){Hog}, {Novikov}, \&
  {Polnarev}}]{1995A&A...294..287H}
{Hog}, E., {Novikov}, I.~D., \& {Polnarev}, A.~G. 1995, \aap, 294, 287

\bibitem[{{Honma}(2001)}]{2001PASJ...53..233H}
{Honma}, M. 2001, \pasj, 53, 233

\bibitem[{{Jordi} {et~al.}(2010){Jordi}, {Gebran}, {Carrasco}, {de Bruijne},
  {Voss}, {Fabricius}, {Knude}, {Vallenari}, {Kohley}, \&
  {Mora}}]{2010A&A...523A..48J}
{Jordi}, C., {Gebran}, M., {Carrasco}, J.~M., {et~al.} 2010, \aap, 523, A48

\bibitem[{{Kl\"uter} {et~al.}(2018){Kl\"uter}, {Bastian}, {Demleitner}, \&
  {Wambsganss}}]{2018arXiv180508023K}
{Kl\"uter}, J., {Bastian}, U., {Demleitner}, M., \& {Wambsganss}, J. 2018,
  /aap, 615, L11

\bibitem[{{Lindegren} {et~al.}(2018){Lindegren}, {Hernandez}, {Bombrun},
  {Klioner}, {Bastian}, {Ramos-Lerate}, {de Torres}, {Steidelmuller},
  {Stephenson}, {Hobbs}, {Lammers}, {Biermann}, {Geyer}, {Hilger}, {Michalik},
  {Stampa}, {McMillan}, {Castaneda}, {Clotet}, {Comoretto}, {Davidson},
  {Fabricius}, {Gracia}, {Hambly}, {Hutton}, {Mora}, {Portell}, {van Leeuwen},
  {Abbas}, {Abreu}, {Altmann}, {Andrei}, {Anglada}, {Balaguer-Nunez},
  {Barache}, {Becciani}, {Bertone}, {Bianchi}, {Bouquillon}, {Bourda},
  {Brusemeister}, {Bucciarelli}, {Busonero}, {Buzzi}, {Cancelliere},
  {Carlucci}, {Charlot}, {Cheek}, {Crosta}, {Crowley}, {de Bruijne}, {de
  Felice}, {Drimmel}, {Esquej}, {Fienga}, {Fraile}, {Gai}, {Garralda},
  {Gonzalez-Vidal}, {Guerra}, {Hauser}, {Hofmann}, {Holl}, {Jordan},
  {Lattanzi}, {Lenhardt}, {Liao}, {Licata}, {Lister}, {Loffler}, {Marchant},
  {Martin-Fleitas}, {Messineo}, {Mignard}, {Morbidelli}, {Poggio}, {Riva},
  {Rowell}, {Salguero}, {Sarasso}, {Sciacca}, {Siddiqui}, {Smart}, {Spagna},
  {Steele}, {Taris}, {Torra}, {van Elteren}, {van Reeven}, \&
  {Vecchiato}}]{2018arXiv180409366L}
{Lindegren}, L., {Hernandez}, J., {Bombrun}, A., {et~al.} 2018, ArXiv e-prints
  [\eprint[arXiv]{1804.09366}]

\bibitem[{{Lindegren} {et~al.}(2016){Lindegren}, {Lammers}, {Bastian},
  {Hern{\'a}ndez}, {Klioner}, {Hobbs}, {Bombrun}, {Michalik}, {Ramos-Lerate},
  {Butkevich}, {Comoretto}, {Joliet}, {Holl}, {Hutton}, {Parsons},
  {Steidelm{\"u}ller}, {Abbas}, {Altmann}, {Andrei}, {Anton}, {Bach},
  {Barache}, {Becciani}, {Berthier}, {Bianchi}, {Biermann}, {Bouquillon},
  {Bourda}, {Br{\"u}semeister}, {Bucciarelli}, {Busonero}, {Carlucci},
  {Casta{\~n}eda}, {Charlot}, {Clotet}, {Crosta}, {Davidson}, {de Felice},
  {Drimmel}, {Fabricius}, {Fienga}, {Figueras}, {Fraile}, {Gai}, {Garralda},
  {Geyer}, {Gonz{\'a}lez-Vidal}, {Guerra}, {Hambly}, {Hauser}, {Jordan},
  {Lattanzi}, {Lenhardt}, {Liao}, {L{\"o}ffler}, {McMillan}, {Mignard}, {Mora},
  {Morbidelli}, {Portell}, {Riva}, {Sarasso}, {Serraller}, {Siddiqui}, {Smart},
  {Spagna}, {Stampa}, {Steele}, {Taris}, {Torra}, {van Reeven}, {Vecchiato},
  {Zschocke}, {de Bruijne}, {Gracia}, {Raison}, {Lister}, {Marchant},
  {Messineo}, {Soffel}, {Osorio}, {de Torres}, \&
  {O'Mullane}}]{2016A&A...595A...4L}
{Lindegren}, L., {Lammers}, U., {Bastian}, U., {et~al.} 2016, \aap, 595, A4

\bibitem[{{Luri} {et~al.}(2018){Luri}, {Brown}, {Sarro}, {Arenou},
  {Bailer-Jones}, {Castro-Ginard}, {de Bruijne}, {Prusti}, {Babusiaux}, \&
  {Delgado}}]{2018arXiv180409376L}
{Luri}, X., {Brown}, A.~G.~A., {Sarro}, L.~M., {et~al.} 2018, ArXiv e-prints
  [\eprint[arXiv]{1804.09376}]

\bibitem[{{McGill} {et~al.}(2018){McGill}, {Smith}, {Evans}, {Belokurov}, \&
  {Smart}}]{2018MNRAS.478L..29M}
{McGill}, P., {Smith}, L.~C., {Evans}, N.~W., {Belokurov}, V., \& {Smart},
  R.~L. 2018, \mnras, 478, L29

\bibitem[{{Miralda-Escude}(1996)}]{1996ApJ...470L.113M}
{Miralda-Escude}, J. 1996, \apjl, 470, L113

\bibitem[{{Miyamoto} \& {Yoshii}(1995)}]{1995AJ....110.1427M}
{Miyamoto}, M. \& {Yoshii}, Y. 1995, \aj, 110, 1427

\bibitem[{{Mustill} {et~al.}(2018){Mustill}, {Davies}, \&
  {Lindegren}}]{2018arXiv180511638M}
{Mustill}, A.~J., {Davies}, M.~B., \& {Lindegren}, L. 2018, ArXiv e-prints
  [\eprint[arXiv]{1805.11638}]

\bibitem[{{Paczynski}(1986)}]{1986ApJ...301..503P}
{Paczynski}, B. 1986, \apj, 301, 503

\bibitem[{{Paczynski}(1991)}]{1991ApJ...371L..63P}
{Paczynski}, B. 1991, \apjl, 371, L63

\bibitem[{{Paczynski}(1995)}]{1995AcA....45..345P}
{Paczynski}, B. 1995, \actaa, 45, 345

\bibitem[{{Paczynski}(1996{\natexlab{a}})}]{1996ARA&A..34..419P}
{Paczynski}, B. 1996{\natexlab{a}}, \araa, 34, 419

\bibitem[{{Paczynski}(1996{\natexlab{b}})}]{1996AcA....46..291P}
{Paczynski}, B. 1996{\natexlab{b}}, \actaa, 46, 291

\bibitem[{{Paczy{\'n}ski}(1998)}]{1998ApJ...494L..23P}
{Paczy{\'n}ski}, B. 1998, \apjl, 494, L23

\bibitem[{{Pecaut} \& {Mamajek}(2013)}]{2013ApJS..208....9P}
{Pecaut}, M.~J. \& {Mamajek}, E.~E. 2013, \apjs, 208, 9

\bibitem[{{Proft} {et~al.}(2011){Proft}, {Demleitner}, \&
  {Wambsganss}}]{2011A&A...536A..50P}
{Proft}, S., {Demleitner}, M., \& {Wambsganss}, J. 2011, \aap, 536, A50

\bibitem[{{Sahu} {et~al.}(2017){Sahu}, {Anderson}, {Casertano}, {Bond},
  {Bergeron}, {Nelan}, {Pueyo}, {Brown}, {Bellini}, {Levay}, {Sokol}, {aff1},
  {Dominik}, {Calamida}, {Kains}, \& {Livio}}]{2017Sci...356.1046S}
{Sahu}, K.~C., {Anderson}, J., {Casertano}, S., {et~al.} 2017, Science, 356,
  1046

\bibitem[{{Sahu} {et~al.}(2014){Sahu}, {Bond}, {Anderson}, \&
  {Dominik}}]{2014ApJ...782...89S}
{Sahu}, K.~C., {Bond}, H.~E., {Anderson}, J., \& {Dominik}, M. 2014, \apj, 782,
  89

\bibitem[{{Salaris} \& {Cassisi}(2005)}]{2005essp.book.....S}
{Salaris}, M. \& {Cassisi}, S. 2005, {Evolution of Stars and Stellar
  Populations}, 400

\bibitem[{{Salim} \& {Gould}(2000)}]{2000ApJ...539..241S}
{Salim}, S. \& {Gould}, A. 2000, \apj, 539, 241

\bibitem[{{Taylor}(2005)}]{2005ASPC..347...29T}
{Taylor}, M.~B. 2005, in Astronomical Society of the Pacific Conference Series,
  Vol. 347, Astronomical Data Analysis Software and Systems XIV, ed.
  P.~{Shopbell}, M.~{Britton}, \& R.~{Ebert}, 29

\bibitem[{{The Astropy Collaboration} {et~al.}(2013){The Astropy
  Collaboration}, {Robitaille, Thomas P.}, {Tollerud, Erik J.}, {Greenfield,
  Perry}, {Droettboom, Michael}, {Bray, Erik}, {Aldcroft, Tom}, {Davis, Matt},
  {Ginsburg, Adam}, {Price-Whelan, Adrian M.}, {Kerzendorf, Wolfgang E.},
  {Conley, Alexander}, {Crighton, Neil}, {Barbary, Kyle}, {Muna, Demitri},
  {Ferguson, Henry}, {Grollier, Fr\'ed\'eric}, {Parikh, Madhura M.}, {Nair,
  Prasanth H.}, {G\"unther, Hans M.}, {Deil, Christoph}, {Woillez, Julien},
  {Conseil, Simon}, {Kramer, Roban}, {Turner, James E. H.}, {Singer, Leo},
  {Fox, Ryan}, {Weaver, Benjamin A.}, {Zabalza, Victor}, {Edwards, Zachary I.},
  {Azalee Bostroem, K.}, {Burke, D. J.}, {Casey, Andrew R.}, {Crawford, Steven
  M.}, {Dencheva, Nadia}, {Ely, Justin}, {Jenness, Tim}, {Labrie, Kathleen},
  {Lim, Pey Lian}, {Pierfederici, Francesco}, {Pontzen, Andrew}, {Ptak, Andy},
  {Refsdal, Brian}, {Servillat, Mathieu}, \& {Streicher, Ole}}]{refId0}
{The Astropy Collaboration}, {Robitaille, Thomas P.}, {Tollerud, Erik J.},
  {et~al.} 2013, A\&A, 558, A33

\bibitem[{{Udalski}(2003)}]{2003AcA....53..291U}
{Udalski}, A. 2003, \actaa, 53, 291

\bibitem[{{Udalski} {et~al.}(2015){Udalski}, {Szyma{\'n}ski}, \&
  {Szyma{\'n}ski}}]{2015AcA....65....1U}
{Udalski}, A., {Szyma{\'n}ski}, M.~K., \& {Szyma{\'n}ski}, G. 2015, \actaa, 65,
  1

\bibitem[{{Wambsganss}(2006)}]{2006AnP...518...43W}
{Wambsganss}, J. 2006, Annalen der Physik, 518, 43

\bibitem[{{Zurlo} {et~al.}(2018){Zurlo}, {Gratton}, {Mesa}, {Desidera}, {Enia},
  {Sahu}, {Almenara}, {Kervella}, {Avenhaus}, {Girard}, {Janson}, {Lagadec},
  {Langlois}, {Milli}, {Perrot}, {Schlieder}, {Thalmann}, {Vigan}, {Giro},
  {Gluck}, {Ramos}, \& {Roux}}]{2018arXiv180701318Z}
{Zurlo}, A., {Gratton}, R., {Mesa}, D., {et~al.} 2018, ArXiv e-prints
  [\eprint[arXiv]{1807.01318}]

\end{thebibliography}
\end{document}